\documentclass[a4paper,12pt]{article}
\usepackage{amsfonts}
\usepackage{subfigure}
\usepackage{bm}
\usepackage{amsmath}
\usepackage{amsthm}
\usepackage{amssymb}
\usepackage[singlespacing]{setspace}
\usepackage{url}
\usepackage[mathscr]{eucal}
\usepackage{subfigure}
\usepackage{lscape}
\usepackage{multirow}
\usepackage[debugshow]{tabularx}
\usepackage{threeparttable}
\usepackage{color}
\usepackage[table,xcdraw]{xcolor}
\usepackage{array}
\usepackage{accents}
\usepackage[figuresright]{rotating}
\usepackage[section]{placeins}%
\usepackage{algorithm,algorithmic} 
\usepackage{booktabs}

\usepackage[pdftex,colorlinks=true,hypertexnames=false]{hyperref}
\definecolor{darkblue}{rgb}{0,0,.6}
\hypersetup{citecolor=darkblue,linkcolor=darkblue,urlcolor=darkblue}
\usepackage{graphicx}
\usepackage{orcidlink}
\usepackage[authoryear,sort&compress]{natbib}

\usepackage{geometry}
\providecommand{\U}[1]{\protect\rule{.1in}{.1in}}
\providecommand{\U}[1]{\protect\rule{.1in}{.1in}}
\newcolumntype{?}{!{\vrule width 1pt}}

\newtheorem{assumption}{Assumption}

\newtheorem{theorem}{Theorem}

\newtheorem{remark}{Remark}

\renewcommand{\arraystretch}{1}
\newcommand{\bbe}{\boldsymbol{e}}

\begin{document}

\title{Robust Nonparametric Testing for Structural Changes in Multivariate Volatility via Multiple Quantiles}

\author{Jilin Wu\orcidlink{0000-0002-2221-1947}\thanks{Department of Finance, School of
Economics, and Wang Yanan Institute for Studies in Economics (WISE), Xiamen University, Xiamen, China.}
\and Ruike Wu\orcidlink{0009-0000-3128-6214}\thanks{Correspondence to: School of Economics, Key Laboratory of Mathematical Economics (Shanghai University of Finance and Economics), Ministry of Education, Shanghai University of Finance and Economics, Shanghai, 200433, China. Email: wuruike@mail.shufe.edu.cn.}
\and Zhijie Xiao\orcidlink{0000-0003-3626-8220}\thanks{Department of Economics, Boston College, Boston, the U.S.}
\and Mengxi Zhang\thanks{Zhejiang University of Finance and Economics, Zhejiang, China.}
}
\date{ }
\maketitle

\vspace{-0.5cm}

\vspace{-0.5cm}

\begin{abstract}

We propose an omnibus nonparametric test for structural changes in the multivariate volatility matrix. The test aggregates bounded generalized quantile scores over a range of quantile levels and has a weighted leave-$q$-out $U$-statistic representation. Deleting nearby index pairs renders the centering effect induced by serial dependence asymptotically negligible. All quantities required for implementation, including the variance estimator used for standardization, are constructed under the null, without specifying volatility dynamics under the alternative. The standardized statistic converges to a standard normal distribution. We establish consistency against fixed alternatives that generate a positive integrated quantile-score signal and derive nontrivial local power against smooth departures and increasingly sharp transitions approaching multiple structural breaks. The bounded-score construction avoids the finite fourth- or eighth-moment conditions commonly imposed by least-squares and quasi-likelihood procedures, while aggregation across quantiles uses more distributional information than single-quantile methods. Monte Carlo results show satisfactory size and favorable power under heavy-tailed innovations, with competitive performance under Gaussian innovations. An application to the Fama--French three-factor model provides evidence against stability of the factor covariance matrix over the full sample and several economically relevant subsamples.

\end{abstract}

\textbf{JEL Codes}: C12, C22.

\textbf{Keywords}: Structural change; Multivariate volatility; Multiple quantiles; Heavy-tailed innovations; Nonparametric testing.

\pagebreak

\section{Introduction}
The analysis of structural changes in asset return volatility has attracted considerable attention in the recent econometric literature.  A primary motivation for this line of inquiry is the recognition that time variation in unconditional volatility can generate spurious persistence, appearing as long-range dependence or IGARCH effects \citep[see][]{lamoureux1990persistence, mikosch2004nonstationarities, hafner2010efficient}. 
Beyond their theoretical implications, such changes are also important for empirical work, particularly in volatility forecasting and financial risk management \citep{rapach2008structural, Pettenuzzo201160}. 
Moreover, ignoring structural changes may reduce estimation efficiency \citep{hansen1995rethinking, xu2008adaptive} and distort standard hypothesis testing procedures \citep{kim2002unit, van2005testing, xu2012robustifying, hafner2010efficient}. Therefore, testing for structural changes in unconditional volatility is an important prerequisite for reliable econometric inference.

While there is a rich literature on detecting structural changes in univariate volatility \citep[e.g.,][]{Inclan1994, Chu1995, xu2013powerful, amado2013modelling, wu2018testing, Wu2026a}, the multivariate counterpart remains comparatively sparse.  Early contributions, such as \cite{galeano2007covariance}, employed Likelihood Ratio (LR) and Cumulative Sum (CUSUM) statistics under Gaussian assumptions. Subsequent studies have relaxed these distributional restrictions; for instance, \cite{aue2009break} developed CUSUM and Quadratic Sum (QS) tests robust to mixing dependence, while \cite{jiang2023testing} introduced a generalized Hausman test capable of detecting both abrupt and smooth changes. Other advances include \cite{korkas2017multiple}, who used wavelet-based segmentation, and \cite{steland2020testing} and \cite{li2023detection}, who studied high-dimensional settings via factor models.  However, many existing procedures rely on finite fourth or even eighth moments to establish their asymptotic validity. This requirement can be restrictive for financial time series, which often exhibit severe leptokurtosis or even infinite variance \citep{akgiray1988stable, loretan1994testing, chen2015sign, wang2022hybrid}. Consequently, developing testing procedures that relax moment requirements is important for empirical applications involving heavy-tailed data.

To date, the literature on structural change testing under weak moment conditions remains limited, with only a few notable exceptions.  \citeauthor{Wu2026a} (\citeyear{Wu2026a}, \citeyear{wu2026b}) pioneered the use of Least Absolute Deviation (LAD) regression to construct robust CUSUM and QS tests for univariate and multivariate volatility, respectively. While LAD-based approaches can improve upon least-squares (LS) methods in heavy-tailed environments, they still have two limitations. First, their exclusive reliance on the conditional median may lead to information loss. Since structural changes in volatility can affect different parts of the conditional distribution, restricting inference to the median may overlook relevant distributional information and reduce testing power. Second, these tests are designed mainly for alternatives involving one single abrupt break. This specification may have limited diagnostic power against multiple breaks or smooth structural changes.  As noted by \citet{Hansen2001}, structural shifts induced by policy reforms or technological diffusion may unfold gradually because of agent learning and adjustment costs. At the same time, the continual
arrival of new information and recurrent changes in market conditions may generate several shifts in volatility dynamics, making multiple-break patterns empirically relevant. Procedures tailored to
a single abrupt break may therefore have limited power against gradual changes or multiple structural breaks.

This paper develops a nonparametric framework for detecting structural changes in multivariate volatility under weak moment conditions. The starting point is an auxiliary regression of generalized quantile scores on rescaled time. Under the null hypothesis, the score process is orthogonal to time and the regression function is identically zero. Changes in volatility, by contrast, induce systematic time variation in the scores. This distinction leads naturally to an $L_2$-type statistic that integrates the squared regression function over time and a range of quantile levels. The statistic can be written as a weighted leave-$q$-out $U$-statistic and is sensitive to both gradual variation and sharp transitions approaching single or multiple structural breaks.

A central feature of the construction is that implementation uses only quantities estimated under the null hypothesis. Deleting index pairs within lag $q$ makes the dependence-induced centering term asymptotically negligible, so this term need not be estimated or explicitly corrected. A feasible null-based variance estimator then yields a standard normal limit after standardization. The procedure therefore requires neither prior specification nor estimation of the volatility dynamics under the alternative. The power analysis covers a broad class of departures from volatility constancy. The test is consistent whenever the fixed alternative generates a positive integrated quantile-score signal, without requiring prior knowledge of the number, locations, or temporal profiles of the changes. It also has nontrivial local power against smoothly evolving departures and increasingly sharp transitions converging to multiple structural breaks. Finally, because the quantile scores are bounded, the asymptotic theory does not impose the finite fourth- or eighth-moment conditions commonly required by least-squares and quasi-likelihood procedures. Integration over a range of quantiles also uses information from different parts of the distribution, reducing the information loss associated with single-quantile methods while retaining robustness to heavy-tailed innovations.

The remainder of the paper is organized as follows. Section~2 introduces the econometric framework and develops the weighted leave-$q$-out $U$-statistic. Section~3 establishes the asymptotic null distribution, consistency under fixed alternatives, and local power against smooth and sharp-transition alternatives. Section~4 describes the numerical implementation and tuning-parameter selection procedures and reports the Monte Carlo results. Section~5 presents an empirical application to the Fama--French three factors,
and Section~6 concludes. All proofs and additional simulation results are provided in the Supplementary Material.

Throughout the paper, let $\mathcal I=[\underline\tau,\overline\tau]\subset(0,1)$ be fixed. For a vector $a=(a_1,\ldots,a_n)^\top$, define $\lVert a\rVert_1=\sum_{i=1}^n|a_i|$ and $\lVert a\rVert_2=(\sum_{i=1}^na_i^2)^{1/2}$. For a matrix $A$, $A^{(i,j)}$, $A^\top$, $\operatorname{tr}(A)$, and $\lVert A\rVert_F=\{\operatorname{tr}(A^\top A)\}^{1/2}$ denote
its $(i,j)$th entry, transpose, trace, and Frobenius norm, respectively, while $\operatorname{vech}(A)$ stacks its lower triangular elements into a vector. For $\tau\in\mathcal I$, define $\psi_\tau(z)=\tau-\mathbb I(z\leq0)$ and $\rho_\tau(z)=z\psi_\tau(z)$, where $\mathbb I(\cdot)$ is the indicator function. For a matrix $A$, $\Psi_\tau(A)$ and $\boldsymbol{\rho}_\tau(A)$ apply these functions elementwise. The symbol $\circ$ denotes the Hadamard product, and $\xrightarrow{d}$ and $\xrightarrow{p}$ denote convergence in distribution and probability. Finally, $\mathbf 0$ denotes a conformable zero vector or matrix, and $\lfloor\cdot\rfloor$ is the floor function.

\section{The Model and The Test}
In this section, we formally define the multivariate volatility model, specify the null and alternative hypotheses, and derive a nonparametric test statistic characterized by a U-statistic structure.

\subsection{Model Specification and Hypotheses}
Consider a $d$-dimensional time series $u_{t}=(u_{1t},\ldots,u_{dt})^{\top}$ for $t=1,\ldots,T$.  Without loss of generality, we assume that $u_t$ has zero conditional mean and admits the following multiplicative representation:
\begin{equation}
u_{t}=\Sigma_{t}^{1/2}\varepsilon_{t}, \label{eq: MDGP1}
\end{equation}
where $\Sigma_{t}$ is a $d\times d$ symmetric positive definite volatility matrix, and $\varepsilon_{t}=(\varepsilon_{1t},\ldots,\varepsilon_{dt})^{\top}$ is a strictly stationary mixing error process. In this representation, $\Sigma_{t}$ describes the time-varying component of multivariate volatility, whereas $\varepsilon_{t}$ captures stationary short-run fluctuations and may allow for multivariate GARCH-type dependence. Thus, the source of nonstationarity in $u_t$ is the time variation in $\Sigma_t$, rather than stochastic trends or unit root behavior. This multiplicative framework covers a broad class of locally stationary volatility models and is consistent with the formulations in \cite{hafner2010efficient}, \cite{patilea2013corrected}, \cite{xu2013powerful}, and \cite{jiang2023testing}.

The primary objective of this paper is to test for structural changes in the volatility matrix $\Sigma_{t}$. Accordingly, the null hypothesis is formulated as
\begin{equation}
 \mathbb{H}_{0}: \Sigma_{t} = \Sigma_{0} \quad \text{for all } t=1,\ldots,T,\label{HYPO}
\end{equation}
where $\Sigma_{0}$ is an unknown constant positive definite matrix. Under $\mathbb{H}_{0}$, the unconditional volatility matrix remains unchanged over time, providing a basis for econometric inference and forecasting procedures that rely on a stable second-moment structure. The alternative hypothesis allows the volatility matrix to change over time:
\begin{equation}
 \mathbb{H}_{A}: \Sigma_{t} \neq \Sigma_{0} \quad \text{for some } t=1,\ldots,T. \label{HYPA}
\end{equation}
Following the infill asymptotic framework of \citet{robinson1989nonparametric}, we represent the time-varying volatility matrix under the alternative as $\Sigma_t=\Sigma(t/T)$, where $\Sigma(\cdot):[0,1]\to\mathbb R^{d\times d}$ is deterministic and positive definite. This representation accommodates smooth changes, increasingly sharp transitions that approximate abrupt breaks, and mixed patterns in which different elements of the volatility matrix exhibit different forms of time variation. For example, one element may undergo a sharp transition, while another changes gradually over time.

\subsection{The U-statistic Test}
Standard testing procedures within the LS framework typically require finite fourth moments, and in some cases even finite eighth moments. These moment conditions, together with the sensitivity of LS estimators to outliers, may weaken the reliability of such tests when they are applied to heavy-tailed macroeconomic and financial time series. To address this issue, we propose a robust nonparametric test for structural changes in the volatility matrix $\Sigma_{t}$. By using quantile regression and combining information from multiple quantiles, the proposed approach is able to handle heavy-tailed errors while relaxing the moment requirements commonly imposed by traditional LS-based methods.

For each $\tau\in\mathcal I$, let $G_t(\tau)=\{G_t^{(i,j)}(\tau)\}_{i,j=1}^d$ denote the elementwise quantile matrix of $u_tu_t^\top$, where $G_t^{(i,j)}(\tau):=\inf\{x:F_t^{(i,j)}(x)\geq\tau\}$ and $F_t^{(i,j)}$ is the marginal CDF of $u_{it}u_{jt}$. The DGP in \eqref{eq: MDGP1} admits the representation
\begin{equation}
u_tu_t^\top
=
G_t(\tau)+\bbe_t(\tau),
\label{eq: ut basic}
\end{equation}
where $E\{\Psi_\tau(\bbe_t(\tau))\}=\mathbf0$. Under $\mathbb H_0$, the constancy of $\Sigma_t$ and the stationarity of $\{\varepsilon_t\}$ imply that $G_t(\tau)=G_0(\tau)$ for every $t$. Under $\mathbb H_A$, time variation in $\Sigma_t$ generally induces corresponding variation in $G_t(\tau)$. Structural stability of the volatility matrix can therefore be examined through the time invariance of the quantile matrices.

Define the pooled marginal CDF by $\overline F_T^{(i,j)}(x):=T^{-1}\sum_{t=1}^T F_t^{(i,j)}(x)$, and denote its $\tau$-quantile by $G_{T,*}^{(i,j)}(\tau)$. Let $G_{T,*}(\tau)$ be the symmetric matrix formed by these elementwise quantiles. Under $\mathbb H_0$,
$G_{T,*}(\tau)=G_0(\tau)$.  Set $M_{t,T,*}(\tau):=\operatorname{vech}\!\left[\Psi_\tau\{u_tu_t^\top-G_{T,*}(\tau)\}\right]$ and $m_T(t/T,\tau):=E\{M_{t,T,*}(\tau)\}$. The score process can then be written as
\begin{equation}
M_{t,T,*}(\tau)
=
m_T(t/T,\tau)+v_{t,T}(\tau),
\qquad
E\{v_{t,T}(\tau)\}=\mathbf0.
\label{eq: EQV}
\end{equation}
The entry of $m_T(t/T,\tau)$ corresponding to $(i,j)$ is $\tau-F_t^{(i,j)}
\{G_{T,*}^{(i,j)}(\tau)\}$. Hence, $T^{-1}\sum_{t=1}^Tm_T(t/T,\tau)=\mathbf0$ for every $\tau\in\mathcal I$. Under $\mathbb H_0$, the mean score vanishes for every $t$ and $\tau$; under $\mathbb H_A$, it may vary with
$t/T$. The pooled quantile matrix is estimated by
\begin{equation}
\widehat G(\tau)
=
\arg\min_{G\in\mathbb R^{d\times d}}
\sum_{t=1}^T
\boldsymbol{\rho}_\tau(u_tu_t^\top-G).
\label{eq: QR}
\end{equation}
Thus, $\widehat G(\tau)$ estimates $G_0(\tau)$ under $\mathbb H_0$ and $G_{T,*}(\tau)$ under $\mathbb H_A$. Let $\widehat\bbe_t(\tau):=
u_tu_t^\top-\widehat G(\tau)$. A local estimator of the mean score is
\begin{equation}
\widehat m(r,\tau)
=
\frac{
\sum_{s=1}^T
k\left\{(s-\lfloor Tr\rfloor)/(Th)\right\}
\operatorname{vech}\{\Psi_\tau(\widehat\bbe_s(\tau))\}
}{
\sum_{s=1}^T
k\left\{(s-\lfloor Tr\rfloor)/(Th)\right\}
}.
\label{eq: NW}
\end{equation}
Here, $k(\cdot)$ is a kernel function and $h$ is a bandwidth parameter.\footnote{Local linear estimation can also be used to
reduce boundary bias, with qualitatively similar asymptotic results
for the present testing problem.} Aggregating the squared score deviations over time and quantile levels gives
\begin{equation}
\widetilde\lambda_T
=
\int_0^1\int_{\mathcal I}
\|\widehat m(r,\tau)\|_2^2
\phi(r)w(\tau)\,d\tau\,dr.
\label{L2TSE}
\end{equation}
Restricting the quantile integration to the prespecified interior interval $\mathcal I$ avoids extreme quantiles, for which estimation may be unstable because of the limited number of effective tail observations, and facilitates uniform asymptotic analysis. The interval can nevertheless be chosen sufficiently wide to retain information from different parts of the distribution. The weighting functions $\phi(r)$ and $w(\tau)$ determine the relative contributions of different time points and quantile levels, respectively.

The behavior of $\widetilde\lambda_T$ is determined by the empirical score process $\{\Psi_\tau(\widehat\bbe_t(\tau))\}$. Under $\mathbb H_0$, $G_{T,*}(\tau)=G_0(\tau)$ and $\widehat G(\tau)$ consistently estimates this common quantile matrix. The empirical scores are therefore asymptotically centered, and $\widehat m(r,\tau)$ converges to $\mathbf0$ for each $(r,\tau)$. In this case, $\widetilde\lambda_T$ is governed by stochastic fluctuations of the centered score process. Under $\mathbb H_A$, $\widehat G(\tau)$ estimates the pooled target $G_{T,*}(\tau)$, which generally differs from the time-specific matrices $G_t(\tau)$. The resulting score process has the time-varying mean $m_T(t/T,\tau)$, which is locally captured by $\widehat m(r,\tau)$. The statistic accumulates the squared magnitude of this deterministic component over the time and quantile domains, thereby providing power against structural changes in $\Sigma_t$.

\begin{remark}
Kernel smoothing near the endpoints generally involves asymmetric effective windows and may therefore generate boundary effects. These effects do not alter the first-order validity of the proposed test.
Under $\mathbb H_0$, the target function $m_T(r,\tau)$ is identically zero on $[0,1]\times\mathcal I$, so the usual boundary bias arising from the local approximation of a time-varying regression function vanishes. Moreover, any remaining boundary contribution is confined to neighborhoods of total length $O(h)$ and is asymptotically negligible after integration under the bandwidth conditions imposed below. Consequently, neither boundary trimming nor an explicit boundary correction is required for null inference. Under the alternatives, however, changes occurring close to the sample endpoints are supported by fewer effective observations and may therefore be associated with
some loss of power.
\end{remark}

Although the statistic in \eqref{L2TSE} involves the weighting function $\phi(\cdot)$, the choice of $\phi(\cdot)$ mainly affects the finite-sample weighting scheme.  Under the regularity conditions stated below, it does not change the limiting null distribution or the consistency of the test.  For analytical convenience, we set $\phi(r)=\left\{(Th)^{-1}\sum_{t=1}^{T}k\left((t-\lfloor Tr\rfloor)/(Th)\right)\right\}^{2}$.  Substituting the expression of $\widehat m(r,\tau)$ in \eqref{eq: NW} into \eqref{L2TSE}, the statistic $\widetilde{\lambda}_{T}$ can be rewritten as
\begin{equation}
\widetilde{\lambda}_{T}
=
\frac{1}{T^{2}h}
\sum_{t=1}^{T}
\sum_{s=1}^{T}
c_{t,s}
\int_{\mathcal I}
w(\tau)
\widehat{M}_{t}(\tau)^{\top}
\widehat{M}_{s}(\tau)
d\tau ,
\label{conv}
\end{equation}
where $\widehat{M}_{t}(\tau)=\operatorname{vech}\{\Psi_{\tau}(\widehat{\bbe}_{t}(\tau))\}$ and  $c_{t,s}=h^{-1}\int_{0}^{1}k\left((t-\lfloor Tr\rfloor)/(Th)\right)k\left((s-\lfloor Tr\rfloor)/(Th)\right)dr$. 
As $T\to\infty$, $c_{t,s}$ is asymptotically equivalent to the two-fold convolution kernel 
$k^{(2)}\left((t-s)/(Th)\right):=\int_{-1}^{1}k\left(u-(t-s)/(Th)\right)k(u)\,du$. As pointed out by \cite{cai2015testing}, it is not necessary to use the convolution kernel directly.  Replacing $k^{(2)}\{(t-s)/(Th)\}$ with the standard kernel $k_{t,s}:=k\left((t-s)/(Th)\right)$ preserves the local weighting structure and does not alter the limiting null distribution or consistency of the test.

In the time-series setting, serial dependence in $\{\Psi_{\tau}(\widehat{\bbe}_{t}(\tau))\}$ may generate a non-negligible centering bias in the $U$-statistic, so that the statistic is not automatically centered at zero under the null. To remove this nuisance component, we adopt the ``leave-$q$-out'' strategy of \cite{juhl2013nonparametric} and \cite{wu2018testing}.  Specifically, we exclude diagonal and near-diagonal terms from the summation and define the test statistic as
\begin{equation}
\widetilde{\lambda}_{BT}
=
\frac{1}{Th^{1/2}}
\sum_{t=1}^{T}
\sum_{s \notin B(t,q)}
k_{t,s}
\int_{\mathcal I}
w(\tau)
\widehat{M}_t(\tau)^{\top}
\widehat{M}_s(\tau)
d\tau,
\label{eq: U statistic}
\end{equation}
where $B(t,q)=\{s:|t-s|\leq q\}$ is a truncation window and $q=q(T)$ satisfies $q\to\infty$ as $T\to\infty$. 
The leave-$q$-out construction retains only pairs $(t,s)$ that are separated by more than $q$ observations. 
Under the mixing conditions imposed on the error process, the contribution of such distant pairs is asymptotically centered under $\mathbb{H}_{0}$. Under $\mathbb{H}_{A}$, by contrast, structural changes in $\Sigma_t$ induce systematic departures in the estimated quantile scores, and these departures are accumulated by $\widetilde{\lambda}_{BT}$. As the sample size increases, this accumulated signal dominates the stochastic fluctuation under suitable regularity conditions, leading to consistency against a broad class of structural changes.  The test is also computationally convenient, as it only requires residuals computed under the null and avoids estimating the unspecified alternative model.

\begin{remark}
\label{remark 2}
The terms excluded by the leave-$q$-out construction in \eqref{eq: U statistic} can be written as
\begin{equation}
\frac{1}{\sqrt h}
\sum_{\ell=-q}^{q}
k\!\left(\frac{\ell}{Th}\right)
\operatorname{tr}\!\left\{
\frac{1}{T}
\sum_{t=\max(1,1+\ell)}^{\min(T,T+\ell)}
\int_{\mathcal I}
w(\tau)
\widehat M_t(\tau)
\widehat M_{t-\ell}(\tau)^\top
\,d\tau
\right\}.
\label{LRVE}
\end{equation}
Under $\mathbb H_0$, the regularity conditions imposed below imply that the leading term of \eqref{LRVE} is
\[
\frac{k(0)}{\sqrt h}
\operatorname{tr}\!\left\{
\int_{\mathcal I}
w(\tau)\Theta(\tau,\tau)\,d\tau
\right\},
\]
where $\Theta(\tau_1,\tau_2)$ is the long-run cross-covariance matrix (LRCM) of the quantile-score process defined in
\eqref{eq:Theta0}. Thus, the near-diagonal pairs in $B(t,q)$ capture the serial-dependence component responsible for the centering bias under the null. Removing these pairs eliminates the need to estimate this component for bias correction, although the LRCM is still estimated for variance normalization. In this respect, the truncation lag $q$ plays a role analogous to the bandwidth in HAC estimation, so data-driven HAC bandwidth rules provide a natural guideline for its selection.
\end{remark}

To obtain a feasible test statistic with a pivotal limiting distribution, we standardize the leave-$q$-out statistic
$\widetilde\lambda_{BT}$. The leave-$q$-out construction removes the centering bias induced by serial dependence under the null, whereas the scale of $\widetilde\lambda_{BT}$ depends on the long-run variance of the quantile-score process. We therefore define
\begin{equation}
\widehat U_T
=
\frac{\widetilde\lambda_{BT}}
{\sqrt{\widehat V_{BT}}},
\label{eq:UT}
\end{equation}
where $\widehat V_{BT}$ is the following feasible estimator of the asymptotic variance of $\widetilde\lambda_{BT}$:
\begin{equation}
\widehat V_{BT}
=
\frac{2}{T^2h}
\sum_{t=1}^T
\sum_{s\notin B(t,q)}
k_{t,s}^2
\int_{\mathcal I}\int_{\mathcal I}
w(\tau_1)w(\tau_2)
\operatorname{tr}\!\left\{
\widehat\Theta(\tau_1,\tau_2)
\widehat\Theta(\tau_2,\tau_1)
\right\}
\,d\tau_1\,d\tau_2.
\label{eq:Vhat}
\end{equation}
Here, $\widehat\Theta(\tau_1,\tau_2)$ is the Bartlett HAC estimator of the LRCM of the quantile-score
process, defined by
\begin{equation}
\widehat\Theta(\tau_1,\tau_2)
=
\sum_{\ell=-q+1}^{q-1}
K_0\!\left(\frac{\ell}{q}\right)
\frac{1}{T}
\sum_{t=\max(1,1+\ell)}^{\min(T,T+\ell)}
\widehat M_t^c(\tau_1)
\widehat M_{t-\ell}^c(\tau_2)^\top,
\label{eq:lrcov}
\end{equation}
where $\widehat M_t^c(\tau)=\widehat M_t(\tau)-T^{-1}\sum_{r=1}^T\widehat M_r(\tau)$, $K_0(x)=(1-|x|)\mathbb I\{|x|\leq1\}$ is the Bartlett lag window,
and $q$ is the same truncation lag as in the leave-$q$-out statistic. The sample centering in $\widehat M_t^c(\tau)$ does not affect the asymptotic null limit but may improve the finite-sample performance of the HAC estimator. Under $\mathbb H_0$, $\widehat\Theta(\tau_1,\tau_2)$ consistently estimates
\begin{equation}
\Theta(\tau_1,\tau_2)
:=
\sum_{\ell=-\infty}^{\infty}
E\!\left[
\operatorname{vech}\!\left\{
\Psi_{\tau_1}\bigl(\bbe_t(\tau_1)\bigr)
\right\}
\operatorname{vech}\!\left\{
\Psi_{\tau_2}\bigl(\bbe_{t-\ell}(\tau_2)\bigr)
\right\}^{\top}
\right],
\label{eq:Theta0}
\end{equation}
the LRCM of the quantile-score process. Moreover, since $(T^2h)^{-1}\sum_{t=1}^T\sum_{s\notin B(t,q)}k_{t,s}^2\to\int_{-1}^1k^2(u)\,du$, the asymptotic variance of $\widetilde\lambda_{BT}$ is
\begin{equation}
V
:=
2\int_{-1}^{1}k^2(u)\,du
\int_{\mathcal I}\int_{\mathcal I}
w(\tau_1)w(\tau_2)
\operatorname{tr}\!\left\{
\Theta(\tau_1,\tau_2)
\Theta(\tau_2,\tau_1)
\right\}
\,d\tau_1\,d\tau_2.
\label{eq:V}
\end{equation}
Consequently, $\widehat V_{BT}\xrightarrow{p}V$ under $\mathbb H_0$. We assume $V>0$ to rule out degenerate quantile-score processes. The resulting test is fully feasible because all its components are computed from the estimated residuals $\{\widehat{\bbe}_t(\tau)\}$ under the null, without imposing any informational requirements on volatility under the alternative hypotheses 

\section{Asymptotic Theory}
\label{sec: asym}

This section provides the asymptotic  distribution of our test statistic under the null and investigates its asymptotic power property under the alternatives.

\subsection{Asymptotic null distribution}

To establish the asymptotic distribution of the test statistic $\widehat{U}_{T}$, we make the following assumptions.

\begin{assumption}
The volatility matrix function $\Sigma(\cdot): [0,1] \to \mathbb{R}^{d \times d}$ is symmetric, positive definite, and twice continuously differentiable.
\label{assum: deter Sigma}
\end{assumption}

\begin{assumption}
\label{assum: clt}
The innovation process $\{\varepsilon_t\}$ is strictly stationary and geometrically $\beta$-mixing. Specifically, there exist
constants $C_\beta<\infty$ and $\rho_\beta\in(0,1)$ such that $\beta(\ell)\leq C_\beta\rho_\beta^\ell$ for every $\ell\geq1$.
\end{assumption}

\begin{assumption}
\label{assum: cdf}
For each $1\leq j\leq i\leq d$, the following conditions hold.

\begin{itemize}
\item[(a)] The marginal CDF admits the representation $F_t^{(i,j)}(x)=F^{(i,j)}(t/T,x)$. Let $G^{(i,j)}(r,\tau)$ denote the $\tau$-quantile of $F^{(i,j)}(r,\cdot)$, so that $G_t^{(i,j)}(\tau)=G^{(i,j)}(t/T,\tau)$. There exists a fixed compact interval $\mathcal X^{(i,j)}$ containing $\{G^{(i,j)}(r,\tau):(r,\tau)\in[0,1]\times\mathcal I\}$ in its interior. The function $F^{(i,j)}(r,x)$ is twice continuously differentiable on
$[0,1]\times\mathcal X^{(i,j)}$, and its density $f^{(i,j)}(r,x):=\partial_xF^{(i,j)}(r,x)$ satisfies $c\leq f^{(i,j)}(r,x)\leq C$ for some $0<c<C<\infty$, uniformly over $(r,x)\in[0,1]\times\mathcal X^{(i,j)}$.

\item[(b)] The conditional CDF $F_t^{(i,j)}(\cdot\mid\mathcal F_{t-1})$ admits a density $f_t^{(i,j)}(\cdot\mid\mathcal F_{t-1})$ that is almost surely continuously differentiable on $\mathcal X^{(i,j)}$. For some $0<c<C<\infty$, $c\leq f_t^{(i,j)}(x\mid\mathcal F_{t-1})\leq C$ and $|\partial_xf_t^{(i,j)}(x\mid\mathcal F_{t-1})|\leq C$ almost surely, uniformly over $t$ and $x\in\mathcal X^{(i,j)}$.
\end{itemize}
\end{assumption}

\begin{assumption}
\label{assum kernel}
(a) The kernel $k(\cdot)$ is a bounded, symmetric, and Lipschitz continuous probability density supported on $[-1,1]$.
(b) The bandwidth $h$ and truncation lag $q$ satisfy $h\to0$, $q\to\infty$, $\log(1/h)=o(q)$, $q/(Th)\to0$, and $q/\sqrt T\to0$.
(c) The weight $w(\tau)$ is deterministic and satisfies $0<\underline w\leq w(\tau)\leq\overline w<\infty$ uniformly over $\tau\in\mathcal I$.
\end{assumption}

Assumption~\ref{assum: deter Sigma} imposes smoothness on the volatility path $\Sigma(\cdot)$, as required for the asymptotic expansion of the test statistic. Although exact step functions are not directly covered by the twice-differentiability condition, they can be obtained as limiting cases of increasingly sharp smooth
transitions. Specifically, for a transition function satisfying $\Pi_G(z,\tau)\to\mathbf 0$ as $z\to-\infty$ and $\Pi_G(z,\tau)\to A_G(\tau)$ as $z\to\infty$, uniformly over $\tau\in\mathcal I$, we have $\Pi_G\!\left(\frac{r-r_0}{\chi_T},\tau\right)
\longrightarrow A_G(\tau)\mathbb I\{r>r_0\}$ for every $r\neq r_0$ as $\chi_T\to0$. Theorem~\ref{theorem 3}(b) extends this construction to multiple transition points and formalizes the resulting local approximation to multiple breaks in the quantile matrix. The Monte Carlo evidence indicates that the proposed test has strong finite-sample power against abrupt breaks and performs particularly favorably under multiple-break alternatives.

Assumption~\ref{assum: clt} specifies the temporal dependence of the innovation process. Since the score process $\{\Psi_\tau(\bbe_t(\tau))\}$ is obtained through measurable transformations of $\{\varepsilon_t\}$, its $\beta$-mixing
coefficients are bounded by those of the innovation process. Geometric decay therefore ensures sufficiently weak long-range dependence for the martingale approximation and asymptotic normality
of the weighted quadratic statistic, as well as for consistent estimation of its LRCM.

Assumption~\ref{assum: cdf} imposes uniform regularity conditions on the marginal and conditional distributions near the quantiles of interest. Part~(a) introduces a smooth continuous-time representation
of the marginal distribution and its associated quantile path $G^{(i,j)}(r,\tau)$. The density lower bound ensures local identification and, together with the implicit function theorem, implies that $G^{(i,j)}(r,\tau)$ is jointly twice continuously differentiable in $(r,\tau)$. Part~(b) provides the conditional smoothness and density bounds needed to accommodate serial dependence, justify uniform conditional expansions of the quantile objective, and control the conditional moments of local score perturbations. Under $\mathbb H_0$, the marginal distribution is time-invariant, whereas the conditional distribution may still depend on $\mathcal F_{t-1}$; the conditional regularity conditions are therefore required to hold uniformly over time.

Assumption~\ref{assum kernel}(a) imposes standard regularity conditions on the smoothing kernel. Assumption~\ref{assum kernel}(b) specifies joint rate restrictions on the bandwidth $h$ and the truncation lag $q$. The bandwidth determines the effective smoothing window, whereas $q$ controls both the width of the deleted band in the leave-$q$-out statistic and the truncation order of the LRCM estimator. Specifically, $\log(1/h)=o(q)$ ensures that the dependence remaining beyond lag $q$ is negligible relative to the normalization of the statistic under geometric absolute regularity; $q/(Th)\to0$ makes the deleted band asymptotically negligible
relative to the effective kernel window; and $q/\sqrt T\to0$ controls the effects of boundary terms, martingale-approximation remainders, and estimation of the null quantile matrix. Together with Assumptions~\ref{assum: clt} and \ref{assum: cdf}, these rate conditions support the stochastic approximations and consistency of the feasible LRCM estimator under $\mathbb H_0$.

Assumption~\ref{assum kernel}(c) requires the quantile weight to be positive and uniformly bounded, ensuring that no quantile region in $\mathcal I$ is asymptotically excluded. The uniform weight
$w(\tau)=1$ yields an omnibus statistic, the V-shaped weight $w(\tau)=\max\{\tau,1-\tau\}$ places greater emphasis on the tails, and the quadratic weight $w(\tau)=4\tau(1-\tau)$ emphasizes the
center of the distribution. The numerical approximation of the quantile integral is described in
Section~\ref{subsec:numerical_implementation}.

\begin{remark}

Unlike conventional least-squares- and quasi-likelihood-based procedures, the proposed test does not impose finite fourth- or eighth-moment conditions. Its asymptotic validity is instead established under the boundedness of the quantile scores and the weak-dependence conditions specified above. Consequently, the test accommodates multivariate volatility models with heavy-tailed innovations, including cases in which the higher-order moments required by conventional procedures do not exist. This robustness is attributable to the boundedness of the quantile scores, rather than to the use of any particular quantile. Aggregating the scores over $\mathcal{I}$ incorporates information from multiple regions of the conditional distribution without imposing a parametric likelihood. Although such aggregation does not confer likelihood efficiency in a formal sense, it can mitigate the loss of information inherent in procedures based on a single quantile. Consistent with this interpretation, the Monte Carlo results indicate that the proposed test delivers higher power under heavy-tailed innovations while exhibiting comparable performance under Gaussian innovations.
\end{remark}

The following theorem shows that the statistic $\widehat{U}_{T}$ under the null hypothesis converges in distribution to the standard normal distribution.

\begin{theorem}
\label{theorem 1}
Suppose Assumptions \ref{assum: deter Sigma}--\ref{assum kernel}
hold. Under $\mathbb H_0$, $\widehat U_T\xrightarrow{d}\mathcal N(0,1)$ as $T\to\infty$.
\end{theorem}
The proof of Theorem~1 proceeds in two steps. First, we establish the asymptotic equivalence between the feasible statistic and its infeasible counterpart constructed from the latent quantile scores. The leave-$q$-out scheme excludes index pairs over which serial dependence remains nonnegligible, thereby reducing the resulting centering term to $o_p(1)$. Second, we derive the limiting distribution of the infeasible
statistic. Because it is a degenerate, time-weighted quadratic form in a serially dependent score process, classical $U$-statistic limit theory for independent observations \citep{hall1984central} is not
directly applicable. Under geometric absolute regularity, the score process admits a martingale approximation, which reduces the infeasible statistic, up to an $o_p(1)$ remainder, to a quadratic form
in martingale differences. We then verify the conditional Lindeberg condition and show that the predictable quadratic variation converges to a limit determined by the long-run covariance matrix of the original score process. The martingale central limit theorem of \citet{hall1980martingale} consequently yields asymptotic normality with this limiting variance. Since the feasible variance estimator is consistent for the same limit, Slutsky's theorem gives the standard normal limit stated in Theorem~1.

\subsection{Asymptotic Power}

We next study the asymptotic power of $\widehat U_T$ under fixed and Pitman-type local alternatives. We first consider fixed alternatives. Define the integrated population signal by
\begin{equation}
\Lambda_{A,T}
:=
\frac1T\sum_{t=1}^T
\int_{\mathcal I}
w(\tau)
\|m_T(t/T,\tau)\|_2^2\,d\tau.
\label{eq:fixed-population-signal}
\end{equation}
The condition $\Lambda_{A,T}\to\Lambda_A>0$ requires the time variation in the pooled quantile-score mean to remain nonvanishing asymptotically. The following theorem establishes consistency under this condition.

\begin{theorem}
\label{theorem 2}
Suppose Assumptions~\ref{assum: deter Sigma}--\ref{assum kernel} hold.  Under $\mathbb H_A$, suppose that $\Lambda_{A,T}\to\Lambda_A>0$. Then, for any positive nonstochastic sequence $\{C_T\}$ satisfying $C_T=o(Th^{1/2}/q)$,
\[
\Pr(\widehat U_T>C_T)\to1.
\]
\end{theorem}

The condition $\Lambda_A>0$ excludes alternatives that modify $\Sigma_t$ without inducing a nonzero deviation in the considered quantile-score process. Under $\mathbb H_A$ and $\Lambda_A>0$, the Supplementary Material shows that the numerator of $\widehat U_T$ contains a positive leading component of order $Th^{1/2}$. Meanwhile, the long-run variance estimator satisfies $\widehat V_{BT}=O_p(q^2)$, since the deviation under the alternative accumulates over the truncation window. Therefore, the standardized statistic diverges to infinity in probability at the rate
$Th^{1/2}/q$. Consequently, Theorem~\ref{theorem 2} establishes the consistency of the proposed test against fixed alternatives satisfying $\Lambda_A>0$. Together with the asymptotic standard normality of $\widehat U_T$ under $\mathbb H_0$, this result motivates the one-sided testing procedure: the null hypothesis is rejected at level $\alpha$ whenever $\widehat U_T>z_\alpha$, where $z_\alpha$ denotes the upper $\alpha$-quantile of $\mathcal N(0,1)$. The procedure is fully data-driven and does not require additional boundary trimming or prior specification of the alternative.

The preceding representation establishes the equivalence between volatility constancy and time invariance of the quantile matrix. For the local asymptotic analysis, it is also necessary to relate the magnitudes of departures under the two parametrizations. Let $\mathcal G(\Sigma,\tau)$ denote the elementwise quantile matrix of
$uu^\top$ when $u=\Sigma^{1/2}\varepsilon_t$, so that $G_t(\tau)=\mathcal G(\Sigma_t,\tau)$ and $G_0(\tau)=\mathcal G(\Sigma_0,\tau)$. The smoothness and density
conditions in Assumptions~\ref{assum: deter Sigma} and \ref{assum: cdf}(a), together with the local identification condition, imply the following local norm equivalence: there exist constants
$0<c_G\leq C_G<\infty$, independent of $T$, such that
\begin{equation}
c_G\sup_{t\leq T}\|\Sigma_t-\Sigma_0\|_F
\leq
\sup_{t\leq T,\,\tau\in\mathcal I}
\|G_t(\tau)-G_0(\tau)\|_F
\leq
C_G\sup_{t\leq T}\|\Sigma_t-\Sigma_0\|_F
\label{eq:Sigma-G-local-equivalence}
\end{equation}
for volatility paths in a sufficiently small neighborhood of $\Sigma_0$. Consequently, local departures in the volatility and quantile parametrizations have the same asymptotic order. The differentiability of $\mathcal G$ further implies that the temporal smoothness of $\Sigma(t/T)$ is inherited by the associated quantile path. Accordingly, the Pitman alternatives are formulated directly in terms of $G_t(\tau)-G_0(\tau)$, covering both smoothly evolving
departures and transition paths converging to multiple structural breaks.

\noindent\textbf{Case I: Local smooth alternatives.}
Consider
\begin{equation}
\mathbb H_{LA}^{1}:\quad
G_t(\tau)
=
G_0(\tau)+\zeta_T\Delta_G(t/T,\tau),
\qquad
1\leq t\leq T,\quad \tau\in\mathcal I,
\label{local1}
\end{equation}
where $\zeta_T\to0$ and $\Delta_G:[0,1]\times\mathcal I\to\mathbb R^{d\times d}$ is symmetric, twice continuously differentiable in $r$, and continuously differentiable in $\tau$. The functions $\Delta_G$, $\partial_r\Delta_G$, $\partial_r^2\Delta_G$, and $\partial_\tau\Delta_G$ are uniformly bounded on
$[0,1]\times\mathcal I$. The function $\Delta_G(r,\tau)$ determines the direction and temporal profile of the departure, whereas
$\zeta_T$ controls its magnitude. Thus, $\mathbb H_{LA}^{1}$ approaches the null while preserving the time profile of the change.
Because the null allows any time-invariant quantile matrix, the alternative is nontrivial only if $\Delta_G(r,\tau)$ varies with $r$ on a set of positive measure.

\noindent\textbf{Case II: Local sharp-transition alternatives.}
To accommodate changes occurring over shrinking transition regions
and converging to multiple breaks, consider
\begin{equation}
\mathbb H_{LA}^{2}:\quad
G_t(\tau)
=
G_0(\tau)
+
\xi_T\sum_{j=1}^Jc_j
\Pi_{G,j}\!\left(
\frac{t/T-r_j}{\chi_T},\tau
\right),
\qquad
1\leq t\leq T,\quad \tau\in\mathcal I,
\label{local2}
\end{equation}
where $\xi_T\to0$ controls the common magnitude of the departures, $\chi_T\to0$ controls the transition width, $J<\infty$ is fixed, and $c_j\in\mathbb R$ determines the relative magnitude and direction of
the $j$th transition. The transition locations $0<r_1<\cdots<r_J<1$ satisfy $\min_{0\leq j\leq J}(r_{j+1}-r_j)\geq c_r>0$, where $r_0:=0$ and $r_{J+1}:=1$.

For each $j$, $\Pi_{G,j}:\mathbb R\times\mathcal I\to\mathbb R^{d\times d}$ is symmetric, twice continuously differentiable in $z$, and continuously
differentiable in $\tau$. The functions $\Pi_{G,j}$, $\partial_z\Pi_{G,j}$, $\partial_z^2\Pi_{G,j}$, and
$\partial_\tau\Pi_{G,j}$ are uniformly bounded on $\mathbb R\times\mathcal I$. Uniformly over $\tau\in\mathcal I$,
suppose that $\Pi_{G,j}(z,\tau)\to\mathbf0$ as $z\to-\infty$ and $\Pi_{G,j}(z,\tau)\to A_{G,j}(\tau)$ as $z\to\infty$, where
$A_{G,j}$ is bounded, measurable, and symmetric. Suppose further that, for some $\mathfrak e_j\in L^1(\mathbb R)\cap L^2(\mathbb R)$, $\sup_{\tau\in\mathcal I}\left\|\Pi_{G,j}(z,\tau)-A_{G,j}(\tau)\mathbb I\{z>0\}
\right\|_F\leq \mathfrak e_j(z)$ . As $\chi_T\to0$,
\begin{equation}
\sum_{j=1}^Jc_j
\Pi_{G,j}\!\left(
\frac{r-r_j}{\chi_T},\tau
\right)
\longrightarrow
H_G(r,\tau)
:=
\sum_{j=1}^J
c_jA_{G,j}(\tau)\mathbb I\{r>r_j\}
\label{eq:sharp-break-direction}
\end{equation}
for every $r\notin\{r_1,\ldots,r_J\}$, uniformly over $\tau\in\mathcal I$. Thus, $\mathbb H_{LA}^{2}$ approaches the null
at rate $\xi_T$, while its normalized time profile approaches the piecewise-constant path $H_G$; $\chi_T$ determines the sharpness of the transitions. Because the null allows any time-invariant quantile
matrix, the alternative is nontrivial only if $H_G(r,\tau)$ varies
with $r$ on a set of positive measure, equivalently, if at least one
$c_jA_{G,j}$ is nonzero on a subset of $\mathcal I$ having positive
Lebesgue measure.

\begin{theorem}
\label{theorem 3}
Suppose Assumptions~\ref{assum: deter Sigma}--\ref{assum kernel} hold and $V>0$. Let $f_0\{G_0(\tau)\}:=\big(f_0^{(i,j)}\{G_0^{(i,j)}(\tau)\}\big)_{i,j=1}^d$ denote the matrix of null marginal densities evaluated at the corresponding elementwise quantiles.
\begin{enumerate}
\item[(a)] Under $\mathbb H_{LA}^{1}$ in \eqref{local1}, let $\zeta_T=T^{-1/2}h^{-1/4}$ and suppose that $q\zeta_T\to0$. Then $\widehat U_T\xrightarrow{d}\mathcal N(\delta_{\mathrm{LA},1},1)$, where $\delta_{\mathrm{LA},1}:=\mu_{\mathrm{LA},1}/\sqrt V$ and
\begin{equation}
\mu_{\mathrm{LA},1}
=
\int_0^1\int_{\mathcal I}w(\tau)
\left\|
\operatorname{vech}\!\left[
f_0\{G_0(\tau)\}\circ
\{\overline\Delta_G(\tau)-\Delta_G(r,\tau)\}
\right]
\right\|_2^2\,d\tau\,dr,
\label{eq:mu-LA1}
\end{equation}
with $\overline\Delta_G(\tau):=
\int_0^1\Delta_G(r,\tau)\,dr$.

\item[(b)] Under $\mathbb H_{LA}^{2}$ in \eqref{local2}, let $\xi_T=T^{-1/2}h^{-1/4}$ and suppose that $h/\chi_T\to0$, and $q\xi_T\to0$. Then
$\widehat U_T\xrightarrow{d} \mathcal N(\delta_{\mathrm{LA},2},1)$, where
$\delta_{\mathrm{LA},2}:=\mu_{\mathrm{LA},2}/\sqrt V$ and
\begin{equation}
\mu_{\mathrm{LA},2}
=
\int_0^1\int_{\mathcal I}w(\tau)
\left\|
\operatorname{vech}\!\left[
f_0\{G_0(\tau)\}\circ
\{\overline H_G(\tau)-H_G(r,\tau)\}
\right]
\right\|_2^2\,d\tau\,dr,
\label{eq:mu-LA2}
\end{equation}
with $\overline H_G(\tau):=\int_0^1H_G(r,\tau)\,dr
=\sum_{j=1}^Jc_j(1-r_j)A_{G,j}(\tau)$.
\end{enumerate}
\end{theorem}

For $j=1,2$, Theorem~\ref{theorem 3} implies
\[
\lim_{T\to\infty}
\Pr\!\left(
\widehat U_T>z_\alpha
\mid\mathbb H_{LA}^{j}
\right)
=
1-\Phi\!\left(z_\alpha-\delta_{\mathrm{LA},j}\right),
\qquad j=1,2,
\]
where $z_\alpha:=\Phi^{-1}(1-\alpha)$. The limiting rejection probability equals $\alpha$ when $\delta_{\mathrm{LA},j}=0$ and exceeds $\alpha$ when $\delta_{\mathrm{LA},j}>0$. Under $\mathbb H_{LA}^{1}$, the deterministic quadratic component balances the stochastic fluctuation when $\zeta_T=T^{-1/2}h^{-1/4}$. The pooled quantile removes the time-average component of the departure, which explains the appearance of $\Delta_G(r,\tau)-\overline\Delta_G(\tau)$ in $\mu_{\mathrm{LA},1}$. Under $\mathbb H_{LA}^{2}$, the transition path converges to $H_G$ as $\chi_T\to0$. Each limiting break persists over a nonvanishing part of the sample, and the quadratic signal is therefore of order $\xi_T^2$. Equating $T\sqrt h\,\xi_T^2$ to one gives the same rate $\xi_T=T^{-1/2}h^{-1/4}$. The constants $c_j$ determine the relative magnitudes and directions of the breaks in $\mu_{\mathrm{LA},2}$.

\section{Monte Carlo Simulation}

\subsection{Numerical Implementation}
\label{subsec:numerical_implementation}

Although the proposed statistic is defined by integration over the quantile index set $\mathcal I$, its numerical implementation uses a finite quantile grid. For an equally spaced grid of $n_\tau\geq2$ points, let
$\Delta_{n_\tau}
=(\overline\tau-\underline\tau)/(n_\tau-1)$ and $\tau_{i,n_\tau}=\underline\tau+(i-1)\Delta_{n_\tau}$, $i=1,\ldots,n_\tau$. Write $\mathcal J_{n_\tau}=\{\tau_{1,n_\tau},\ldots,\tau_{n_\tau,n_\tau}\}$, and assign the trapezoidal weights $a_{1,n_\tau}=a_{n_\tau,n_\tau}=\Delta_{n_\tau}/2$ and $a_{i,n_\tau}=\Delta_{n_\tau}$ for
$2\leq i\leq n_\tau-1$. The resulting grid-based numerator is
\begin{equation*}
\widetilde\lambda_{BT,n_\tau}
=
\frac{1}{T\sqrt h}
\sum_{t=1}^T\sum_{s\notin B(t,q)}
k_{t,s}
\sum_{i=1}^{n_\tau}
a_{i,n_\tau}w(\tau_{i,n_\tau})
\widehat M_t(\tau_{i,n_\tau})^\top
\widehat M_s(\tau_{i,n_\tau}).
\end{equation*}

For each fixed sample size $T$, the integrands are bounded and Riemann integrable over the compact quantile interval. Standard trapezoidal quadrature therefore recovers the continuously integrated
statistic as $n_\tau\to\infty$, or equivalently as $\Delta_{n_\tau}\to0$. This is a numerical approximation for each
given sample and is distinct from the asymptotic limit $T\to\infty$ used in the theoretical analysis.

Using the same grid and trapezoidal weights, the feasible variance estimator is computed as
\begin{equation*}
\widehat V_{BT,n_\tau}
=
\frac{2}{T^2h}
\sum_{t=1}^T\sum_{s\notin B(t,q)}
k_{t,s}^2
\sum_{i=1}^{n_\tau}\sum_{j=1}^{n_\tau}
a_{i,n_\tau}a_{j,n_\tau}
w(\tau_{i,n_\tau})w(\tau_{j,n_\tau})
\operatorname{tr}\!\left\{
\widehat\Theta(\tau_{i,n_\tau},\tau_{j,n_\tau})
\widehat\Theta(\tau_{j,n_\tau},\tau_{i,n_\tau})
\right\}.
\end{equation*}
The corresponding self-normalized statistic is $\widehat U_{T,n_\tau}=\widetilde\lambda_{BT,n_\tau}/
\sqrt{\widehat V_{BT,n_\tau}}$. Thus, the numerator and its
self-normalizer are evaluated using the same kernel, leave-$q$-out structure, quantile grid, and quadrature rule.

In all Monte Carlo experiments and the empirical application, we set $\mathcal I=[0.10,0.90]$ and use
$\mathcal J_{17}=\{0.10,0.15,\ldots,0.85,0.90\}$, corresponding to $\Delta_{n_\tau}=0.05$ and $n_\tau=17$. Restricting the grid to this interior interval reduces the instability of extreme-quantile estimation associated with the small number of effective tail observations in finite samples. Unless otherwise stated, we use the
uniform weight $w(\tau)=1$ on $\mathcal J_{17}$.

\subsection{Bandwidth and Truncation Parameter Selection}
\label{subsec:bandwidth_truncation}

The proposed test requires the choice of two tuning parameters: the bandwidth $h$ used in the nonparametric smoothing step and the truncation lag $q$ used in the leave-$q$-out statistic. 
We first discuss the selection of $h$. Since the statistic is constructed from the Nadaraya--Watson estimator in \eqref{eq: NW}, a natural approach is to use cross-validation. For independent observations, leave-one-out cross-validation provides a standard data-driven bandwidth selector. 
In the presence of serial dependence, however, leave-one-out cross-validation may be affected by nearby dependent observations and can lead to inappropriate bandwidth choices. Following \cite{chu1991comparison}, we therefore use a leave-$q_{\mathrm{cv}}$-out cross-validation criterion. For each $t$ and $\tau$, define the leave-$q_{\mathrm{cv}}$-out estimator
\begin{equation}
\widehat m_{\mathrm{cv}}(t/T,\tau)
=
\frac{
\sum_{s:|s-t|>q_{\mathrm{cv}}}
k\left( \frac{s-t}{Th}\right)\widehat{M}_s(\tau)
}{
\sum_{s:|s-t|>q_{\mathrm{cv}}}
k\left( \frac{s-t}{Th}\right)
},
\label{DD}
\end{equation}
where the observations within the window $|s-t|\leq q_{\mathrm{cv}}$ are omitted. The parameter $q_{\mathrm{cv}}$ plays the same role as the truncation lag $q$ in \eqref{eq: U statistic} and is chosen so that $q_{\mathrm{cv}}/(Th)\to0$ as $T\to\infty$.  The bandwidth is then selected by
\begin{equation}
\widehat h_{\mathrm{cv}}
=
\underset{c_1T^{-1/5}\leq h\leq c_2T^{-1/5}}{\arg\min}
\sum_{t=1}^T
\int_{\mathcal I}
w(\tau)
\left\|
\widehat M_t(\tau)
-
\widehat m_{\mathrm{cv},h}(t/T,\tau)
\right\|_2^2
\,d\tau,
\label{DDh}
\end{equation}
where $c_{1}$ and $c_{2}$ are two prespecified constants, and $T^{-1/5}$ is the optimal order of magnitude obtained by minimizing the asymptotic mean integrated squared error of $\widehat m(t/T,\tau)$. The data-driven bandwidth automatically adjusts to the data and leads to a test based on the trade-off between size and power.  Under the null hypothesis, $\widehat{G}(\tau)$ is a consistent estimator for $G_0(\tau)$, and thus the estimator $\widehat m(t/T,\tau)$ is asymptotically close to zero vector over time. Consequently, \eqref{DDh} will select a bandwidth as large as possible. Under the alternative hypothesis, the estimator $\widehat m(t/T,\tau)$ is substantially different from the zero vector, \eqref{DDh} searches for a way of balancing bias and variance and settles down at an optimal bandwidth. Note that the magnitude order $T^{-1/5}$ of candidates satisfies the bandwidth requirement in Assumption \ref{assum kernel}(b). In this paper, we specify $h=1.05^{j-15}T^{-1/5}$, $1\leq j\leq30$, which implies that $c_{1}=0.5051$ and $c_{2}=2.079$.

To determine the truncation parameter $q$, we employ a data-driven procedure that adapts to the serial correlation structure in $\left\{\Psi_{\tau}\left({\bbe}_{t}(\tau)\right) \right\}_{t=1}^{T}$. Following \cite{andrews1991heteroskedasticity}, we utilize a plug-in method based on first-order autocorrelation. Specifically, let $\bar{\epsilon}^{(i,j)}(\tau)$ denote the sample mean of $\epsilon_{t}^{(i,j)}(\tau) = \psi_{\tau}(\bbe_{t}^{(i,j)}(\tau))$.  The average serial dependence is quantified by $\varphi=\frac{2}{d(d+1)}
\sum_{i=1}^{d}\sum_{j=i}^{d}
\int_{\mathcal I}|\rho^{(i,j)}(\tau)|\,d\tau$, where
$$
{\rho}^{(i,j)}(\tau)=\frac{\sum_{t=2}^{T}\left( \epsilon_{t}^{(i,j)}(\tau)-\bar{\epsilon}^{(i,j)}(\tau)\right) \left( \epsilon_{t-1}^{(i,j)}(\tau)-\bar{\epsilon}^{(i,j)}(\tau)\right) }{\sum_{t=1}^{T}\left(\epsilon_{t}^{(i,j)}(\tau)-\bar{\epsilon}^{(i,j)}(\tau)\right)^2}, \quad 1\leq i\leq j\leq d,
$$
and $\mathcal I$ is the quantile interval considered above.

Accordingly, we select the truncation lag as $q=\lfloor \widehat{\varphi}T^{1/3}\rfloor$, where $\widehat{\varphi}$ measures the strength of serial dependence. The rate $T^{1/3}$ follows the conventional Bartlett HAC bandwidth order \citep{andrews1991heteroskedasticity}. When $\widehat{\varphi}=0$, we have $q=0$, so that $B(t,q)=\{t\}$. For the asymptotic analysis, Assumption~\ref{assum kernel}(b) requires $q\to\infty$, $\log(1/h)=o(q)$, $q/(Th)\to0$, and
$q/\sqrt T\to0$. With $h=O(T^{-1/5})$ and $\widehat\varphi=O_p(1)$, the proposed selector satisfies these
conditions, since $q=O_p(T^{1/3})$, $q/(Th)=O_p(T^{-7/15})$, and $q/\sqrt T=O_p(T^{-1/6})$. In implementation, the unobserved score matrix $\epsilon_t(\tau):=\Psi_\tau\{\bbe_t(\tau)\}$ is replaced by $\widehat\epsilon_t(\tau):=\Psi_\tau\{\widehat\bbe_t(\tau)\}$, where $\widehat\bbe_t(\tau)=u_tu_t^\top-\widehat G(\tau)$ is constructed from the quantile residuals under the null.

In addition, since the tuning parameter \( q_{\mathrm{cv}} \) in (\ref{DD}) serves the same purpose as \( q \), we specify \( q_{\mathrm{cv}}=q \) for simplicity. The Monte Carlo simulations show that the procedure for selecting the tuning parameters performs effectively in finite samples. The quantile integration used in both the cross-validation criterion and the final statistic follows the grid and trapezoidal rule described in Subsection~\ref{subsec:numerical_implementation}.

\subsection{Simulation Results}

In this section, we study the performance of the newly proposed test, and compare it with other popular benchmark tests through Monte Carlo experiments. The newly proposed test is denoted as \( \widehat{U}_{T} \). For the benchmarks, we consider the CUSUM and QS tests from \cite{aue2009break}, denoted as \( MS_A \) and \( MQ_A \) respectively; the Hausman tests \( \widehat{\mathcal{D}} \) and \( \widehat{\mathcal{D}}_b \) developed by \cite{jiang2023testing}; and the LAD-based tests \( MS_M^*\), \( MQ_M^*\) and \( T_{CM}^*\) introduced by \cite{wu2026b}. For the test \( \widehat{U}_{T}\), we employ the Bartlett kernel and determine the bandwidth \( h \) as well as the truncation parameter $q$ using the methods described in Subsection~\ref{subsec:bandwidth_truncation}. For the tests \( MS_A \) and \( MQ_A \), we use the Bartlett kernel with the truncation parameter \( q_T = \lfloor T^{1/3} \rfloor \). For the Hausman tests \( \widehat{\mathcal{D}} \) and \( \widehat{\mathcal{D}}_b \), as well as the LAD-based tests \( MS_M^*\), \( MQ_M^*\) and \( T_{CM}^*\), we strictly follow the testing procedures outlined in the simulation sections of \cite{jiang2023testing} and \cite{wu2026b}. Specifically, the critical value for the test \( \widehat{\mathcal{D}} \) is determined based on the asymptotically standard normal distribution $\mathcal{N}\left(  0,1\right)$, whereas the critical value for the test \( \widehat{\mathcal{D}}_b \) is obtained using the wild bootstrap method.

We consider the cases of \( d = 5 \) and $T\in \{250,500,750,1000 \}$ for \( \{ u_t \}_{t=1}^{T} \), with 1000 replications. To investigate size performance of the above-mentioned tests under $\mathbb{H}_{0}$, we consider the following data generating processes (DGPs):
\begin{align}
u_{t}  & =\Sigma_{t}^{1/2}\varepsilon_{t},\quad\varepsilon_{t}=H_{t}^{1/2}\eta_{t},\notag\\
H_{t}  & =I_{5}-A_{0}A_{0}^{\top}-B_{0}B_{0}^{\top}+A_{0}\varepsilon_{t-1}\varepsilon_{t-1}^{\top}A_{0}^{\top}+B_{0}H_{t-1}B_{0}^{\top},\label{SDGP}
\end{align}
where the parameters $A_0$ and $B_0$  are specified as follows:
\[
A_{0}=\left(
\begin{array}
[c]{rrrrr}%
0.30  & 0.02 & -0.03 & -0.02 & 0.02  \\
-0.01 & 0.30 & 0.05  & 0.05  & 0.01  \\
-0.04 & 0.02 & 0.30  & -0.03 & 0.00  \\
-0.03 & 0.08 & 0.04  & 0.30  & -0.01 \\
0.00  & 0.00 & 0.00  & 0.00  & 0.30 \\
\end{array}
\right), B_{0}=\left(
\begin{array}
[c]{rrrrr}%
0.85  & 0.01  & 0.01 & 0.01  & 0.00  \\
-0.01 & 0.85  & 0.02 & -0.02 & -0.01 \\
0.03  & 0.02  & 0.85 & 0.00  & 0.01  \\
0.03  & -0.03 & 0.01 & 0.85  & -0.02 \\
0.00  & 0.00  & 0.00 & -0.06 & 0.85 \\
\end{array}
\right).
\]
The innovation $  \eta_{t}  $ is an i.i.d. random vector in \( \mathbb{R}^5 \), with zero mean and an identity covariance matrix. In order to demonstrate the robustness of the newly developed test against a range of error distributions, especially those with heavy tails, we consider the following cases of error distributions for \( \{ \eta_t \} \): (a) the multivariate normal distribution, $MN$; (b) the standardized multivariate student-t distribution with $3$ degrees of freedom, $MT$; (c) the multivariate skewed student-t distribution with $4$ degrees of freedom and the skewness parameters vector $(-0.8,-0.8,-0.8,-0.8,-0.8)$, $SMT$; (d) the multivariate standardized log normal distribution, $MLN$. The last three disturbances follow heavy-tailed distributions. Without loss of generality, we assume \( \Sigma_t = I_5 \), which implies that \( u_t \) follows a stationary five-dimensional BEKK(1,1) process under $\mathbb{H}_{0}$.

\begin{table}[!htbp] \centering\renewcommand{\arraystretch}{1.15}%
\addtolength{\tabcolsep}{12.0pt}
\caption{Empirical rejection probabilities of the tests under the null.}%
\setlength{\tabcolsep}{3mm}{
\begin{tabular}
[c]{cccccccccc}\toprule
Error & $T$   & $\widehat{\mathcal{D}}$ & $\widehat{\mathcal{D}}_{b}$ & $MS_{A}$ &
$MQ_{A}$ & $MS_M^*$ & $MQ_M^*$ & $T_{CM}^*$ & $\widehat{U}_{T}$ \\ \midrule
$MN$  & 250  & 0.000 & 0.079 & 0.000 & 0.007 & 0.164 & 0.199 & 0.187 & 0.040 \\
      & 500  & 0.008 & 0.072 & 0.016 & 0.098 & 0.136 & 0.169 & 0.163 & 0.051 \\
      & 750  & 0.020 & 0.065 & 0.053 & 0.137 & 0.120 & 0.143 & 0.132 & 0.023 \\
      & 1000 & 0.026 & 0.060 & 0.066 & 0.157 & 0.124 & 0.145 & 0.141 & 0.030 \\
$MT$  & 250  & 0.015 & 0.055 & 0.000 & 0.000 & 0.144 & 0.171 & 0.160 & 0.067 \\
      & 500  & 0.023 & 0.059 & 0.001 & 0.009 & 0.121 & 0.162 & 0.147 & 0.062 \\
      & 750  & 0.033 & 0.061 & 0.005 & 0.041 & 0.135 & 0.161 & 0.154 & 0.055 \\
      & 1000 & 0.029 & 0.054 & 0.008 & 0.039 & 0.111 & 0.124 & 0.118 & 0.050 \\
$SMT$ & 250  & 0.008 & 0.060 & 0.000 & 0.000 & 0.142 & 0.170 & 0.163 & 0.062 \\
      & 500  & 0.012 & 0.067 & 0.005 & 0.032 & 0.128 & 0.150 & 0.148 & 0.063 \\
      & 750  & 0.019 & 0.048 & 0.009 & 0.063 & 0.104 & 0.124 & 0.123 & 0.051 \\
      & 1000 & 0.025 & 0.046 & 0.024 & 0.089 & 0.101 & 0.107 & 0.104 & 0.047 \\
$MLN$ & 250  & 0.027 & 0.073 & 0.000 & 0.001 & 0.127 & 0.140 & 0.140 & 0.071 \\
      & 500  & 0.023 & 0.054 & 0.001 & 0.009 & 0.108 & 0.138 & 0.128 & 0.066 \\
      & 750  & 0.019 & 0.047 & 0.004 & 0.041 & 0.120 & 0.141 & 0.129 & 0.051 \\
      & 1000 & 0.029 & 0.059 & 0.009 & 0.038 & 0.099 & 0.118 & 0.109 & 0.058\\ \bottomrule 
\end{tabular}%
}
\label{Table size}%
\end{table}

Table~\ref{Table size} reports empirical rejection frequencies under the null hypothesis at the 5\% nominal level for $T\in\{250,500,750,1000\}$ and the four innovation distributions. The rejection frequencies of $\widehat U_T$ are generally close to 0.05, although some overrejection is observed in the smaller samples. This distortion becomes less pronounced as $T$ increases. The asymptotic Hausman test $\widehat{\mathcal D}$ tends to underreject across the reported designs. Its bootstrap counterpart, $\widehat{\mathcal D}_b$, is closer to the nominal level in many cases but displays modest overrejection for small $T$. The statistic $MS_A$ also underrejects in most experiments and seldom reaches the nominal level. The size behavior of $MQ_A$ is less uniform: it underrejects in several small-sample designs but overrejects in some other combinations of sample size and innovation distribution. The three LAD-based procedures, $MS_M^*$, $MQ_M^*$, and $T_{CM}^*$, show a different pattern. \citet{wu2026b} report satisfactory size control under moderate volatility persistence, whereas the rejection frequencies in Table~\ref{Table size} remain above 0.05 under the stronger ARCH persistence considered here. The distortion does not disappear quickly with the sample size. Even at $T=1000$, all three procedures continue to overreject across the four innovation distributions. The contrast between the two sets of designs indicates that the finite-sample size of these LAD-based tests can be sensitive to the persistence of the volatility process.

Next, in order to explore the empirical power of the proposed test under alternatives, we continue to utilize the aforementioned DGP, with the exception that \( \Sigma_{t} \) is specified as follows:

DGPP.1-Single structural break:
\begin{equation}
\Sigma_{t}=\left\{
\begin{aligned} \nonumber &\Sigma_0 &t/T\leq 0.5\\ &\Sigma_0+\gamma\Sigma_1\quad &t/T>0.5\\ \end{aligned};\right.
\end{equation}

DGPP.2-Two structural breaks:
\begin{equation}
\Sigma_{t}=\left\{
\begin{aligned} \nonumber &\Sigma_0 &t/T<0.3 \ \text{or} \ t/T> 0.7\\ &\Sigma_0+\gamma\Sigma_1\quad &0.3\leq t/T\leq0.7\end{aligned};\right.
\end{equation}

DGPP.3-Four structural breaks:
\begin{equation}
\Sigma_{t}=\left\{
\begin{aligned} \nonumber &\Sigma_0 &t/T<0.2\ \text{or} \ t/T\geq 0.8\\ &\Sigma_0+0.5\gamma\Sigma_1 \quad &0.2\leq t/T<0.4 \ \text{or}\ 0.6\leq t/T<0.8\\
&\Sigma_0+\gamma\Sigma_1\quad &0.4\leq t/T<0.6
\end{aligned};\right.
\end{equation}

DGPP.4-Smooth structural change with U shape: $\Sigma_{t}=\Sigma_{0}+\gamma\left[1-\exp(-15(t/T-0.5)^2)\right]\Sigma_{1}$;

DGPP.5-Quadratic smooth structural change: $\Sigma_{t}=\Sigma_{0}+\gamma(t/T)^{2}\Sigma_{1}$;

DGPP.6-Oscillating smooth structural change: $\Sigma_{t}=\Sigma_{0}+\gamma\left[  \sin(2\pi t/T)+1\right]  \Sigma_{1}$.

The parameter $\gamma$ is set to $1$ for the time being, later we will treat $\Sigma_{t}$ as a function of $\gamma$ to explore monotonic power of all considered tests.  Finally, we set $\Sigma_0 = I_5$ and $\Sigma_1 = 0.5I_5 + 0.5\mathbf{1}_5\mathbf{1}_5^\top$  where $\mathbf{1}_5$ is the five-dimensional column vector of ones.

\begin{table}[!htbp] \centering\renewcommand{\arraystretch}{1.15}%
\addtolength{\tabcolsep}{12.0pt}
\caption{Empirical rejection probabilities of the tests under DGPP.1.}%
\setlength{\tabcolsep}{3mm}{
\begin{tabular}
[c]{cccccccccc}\toprule
Error & $T$   & $\widehat{\mathcal{D}}$ & $\widehat{\mathcal{D}}_{b}$ & $MS_{A}$ &
$MQ_{A}$ & $MS_M^*$ & $MQ_M^*$ & $T_{CM}^*$ & $\widehat{U}_{T}$ \\ \midrule
$MN$  & 250  & 0.572 & 0.586 & 0.801 & 0.621 & 0.546 & 0.489 & 0.516 & 0.854 \\
      & 500  & 0.964 & 0.945 & 0.995 & 0.951 & 0.960 & 0.910 & 0.945 & 0.998 \\
      & 750  & 0.998 & 0.997 & 1.000 & 0.998 & 0.996 & 0.992 & 0.994 & 1.000 \\
      & 1000 & 0.998 & 0.998 & 1.000 & 1.000 & 1.000 & 0.999 & 1.000 & 1.000 \\
$MT$  & 250  & 0.112 & 0.133 & 0.293 & 0.254 & 0.362 & 0.353 & 0.358 & 0.526 \\
      & 500  & 0.151 & 0.196 & 0.570 & 0.412 & 0.829 & 0.784 & 0.815 & 0.918 \\
      & 750  & 0.254 & 0.290 & 0.763 & 0.569 & 0.969 & 0.940 & 0.959 & 0.997 \\
      & 1000 & 0.318 & 0.320 & 0.835 & 0.718 & 0.998 & 0.989 & 0.998 & 1.000 \\
$SMT$ & 250  & 0.105 & 0.137 & 0.333 & 0.214 & 0.363 & 0.336 & 0.354 & 0.560 \\
      & 500  & 0.322 & 0.340 & 0.657 & 0.496 & 0.878 & 0.815 & 0.859 & 0.940 \\
      & 750  & 0.531 & 0.464 & 0.824 & 0.671 & 0.990 & 0.978 & 0.988 & 0.998 \\
      & 1000 & 0.612 & 0.569 & 0.899 & 0.799 & 0.999 & 0.996 & 0.998 & 1.000 \\
$MLN$ & 250  & 0.042 & 0.085 & 0.543 & 0.395 & 0.852 & 0.806 & 0.820 & 0.994 \\
      & 500  & 0.096 & 0.119 & 0.851 & 0.714 & 0.998 & 0.995 & 0.996 & 1.000 \\
      & 750  & 0.160 & 0.167 & 0.950 & 0.840 & 1.000 & 1.000 & 1.000 & 1.000 \\
      & 1000 & 0.184 & 0.210 & 0.980 & 0.950 & 1.000 & 1.000 & 1.000 & 1.000\\ \bottomrule
\end{tabular}%
}
\label{Table Power DGP1}%
\end{table}

Tables~\ref{Table Power DGP1}--\ref{Table Power DGP6} report rejection frequencies under DGPP.1--6 at the 5\% level. The results are size adjusted for all procedures except $\widehat{\mathcal D}_b$.\footnote{Because $\widehat{\mathcal D}_b$ uses bootstrap critical values, its reported rejection frequencies are not size adjusted.} We begin with DGPP.1, in which a single break occurs at the sample midpoint. Under Gaussian innovations, all tests have substantial power, with $\widehat U_T$ recording the highest rejection frequencies. Under the non-Gaussian distributions, the largest rejection frequencies are generally obtained by $\widehat U_T$ and the three LAD-based tests, $MS_M^*$, $MQ_M^*$, and $T_{CM}^*$. The difference is most pronounced in small samples. For example, when $T=250$ and the innovations follow the $SMT$ distribution, the rejection frequency of $\widehat U_T$ is 0.560, compared with 0.363 for $MS_M^*$, the next highest value. In the same single-break design, $MS_A$ and $MQ_A$ also tend to have higher rejection frequencies than $\widehat{\mathcal D}$ and $\widehat{\mathcal D}_b$.

\begin{table}[!htbp] \centering\renewcommand{\arraystretch}{1.15}%
\addtolength{\tabcolsep}{12.0pt}
\caption{Empirical rejection probabilities of the tests under DGPP.2.}%
\setlength{\tabcolsep}{3mm}{
\begin{tabular}
[c]{cccccccccc}\toprule
Error & $T$   & $\widehat{\mathcal{D}}$ & $\widehat{\mathcal{D}}_{b}$ & $MS_{A}$ &
$MQ_{A}$ & $MS_M^*$ & $MQ_M^*$ & $T_{CM}^*$ & $\widehat{U}_{T}$ \\ \midrule
$MN$  & 250  & 0.249 & 0.264 & 0.039 & 0.021 & 0.068 & 0.089 & 0.088 & 0.773 \\
      & 500  & 0.888 & 0.837 & 0.035 & 0.029 & 0.075 & 0.119 & 0.108 & 0.987 \\
      & 750  & 0.997 & 0.984 & 0.052 & 0.055 & 0.087 & 0.153 & 0.139 & 1.000 \\
      & 1000 & 0.998 & 0.998 & 0.060 & 0.105 & 0.208 & 0.261 & 0.235 & 1.000 \\
$MT$  & 250  & 0.034 & 0.053 & 0.052 & 0.026 & 0.054 & 0.069 & 0.063 & 0.415 \\
      & 500  & 0.093 & 0.130 & 0.047 & 0.017 & 0.077 & 0.111 & 0.094 & 0.836 \\
      & 750  & 0.160 & 0.175 & 0.048 & 0.023 & 0.084 & 0.113 & 0.102 & 0.988 \\
      & 1000 & 0.250 & 0.257 & 0.062 & 0.038 & 0.182 & 0.245 & 0.225 & 1.000 \\
$SMT$ & 250  & 0.034 & 0.045 & 0.037 & 0.011 & 0.055 & 0.076 & 0.068 & 0.511 \\
      & 500  & 0.203 & 0.216 & 0.045 & 0.028 & 0.072 & 0.096 & 0.084 & 0.897 \\
      & 750  & 0.444 & 0.379 & 0.049 & 0.032 & 0.135 & 0.158 & 0.147 & 0.990 \\
      & 1000 & 0.541 & 0.489 & 0.055 & 0.037 & 0.220 & 0.229 & 0.225 & 1.000 \\
$MLN$ & 250  & 0.017 & 0.044 & 0.054 & 0.027 & 0.082 & 0.113 & 0.104 & 0.968 \\
      & 500  & 0.056 & 0.075 & 0.056 & 0.037 & 0.105 & 0.172 & 0.155 & 1.000 \\
      & 750  & 0.085 & 0.092 & 0.065 & 0.051 & 0.154 & 0.238 & 0.214 & 1.000 \\
      & 1000 & 0.128 & 0.147 & 0.067 & 0.084 & 0.373 & 0.418 & 0.406 & 1.000\\ \bottomrule
\end{tabular}%
}
\label{Table Power DGP2}%
\end{table}

\begin{table}[!htbp] \centering\renewcommand{\arraystretch}{1.15}%
\addtolength{\tabcolsep}{12.0pt}
\caption{Empirical rejection probabilities of the tests under DGPP.3.}%
\setlength{\tabcolsep}{3mm}{
\begin{tabular}
[c]{cccccccccc}\toprule
Error & $T$ &   $\widehat{\mathcal{D}}$ & $\widehat{\mathcal{D}}_{b}$ & $MS_{A}$ &
$MQ_{A}$ & $MS_M^*$ & $MQ_M^*$ & $T_{CM}^*$ & $\widehat{U}_{T}$ \\ \midrule
$MN$  & 250  & 0.135 & 0.171 & 0.044 & 0.026 & 0.065 & 0.071 & 0.068 & 0.526 \\
      & 500  & 0.715 & 0.707 & 0.045 & 0.033 & 0.077 & 0.116 & 0.097 & 0.888 \\
      & 750  & 0.964 & 0.939 & 0.053 & 0.066 & 0.064 & 0.131 & 0.106 & 0.998 \\
      & 1000 & 0.993 & 0.981 & 0.055 & 0.110 & 0.125 & 0.243 & 0.221 & 1.000 \\
$MT$  & 250  & 0.032 & 0.040 & 0.046 & 0.030 & 0.049 & 0.058 & 0.054 & 0.223 \\
      & 500  & 0.051 & 0.075 & 0.048 & 0.018 & 0.076 & 0.095 & 0.091 & 0.570 \\
      & 750  & 0.091 & 0.116 & 0.048 & 0.025 & 0.055 & 0.090 & 0.079 & 0.872 \\
      & 1000 & 0.133 & 0.134 & 0.062 & 0.036 & 0.108 & 0.197 & 0.177 & 0.965 \\
$SMT$ & 250  & 0.034 & 0.037 & 0.046 & 0.021 & 0.057 & 0.069 & 0.059 & 0.290 \\
      & 500  & 0.112 & 0.130 & 0.040 & 0.024 & 0.058 & 0.085 & 0.082 & 0.655 \\
      & 750  & 0.270 & 0.226 & 0.049 & 0.027 & 0.093 & 0.138 & 0.124 & 0.893 \\
      & 1000 & 0.351 & 0.312 & 0.047 & 0.028 & 0.141 & 0.209 & 0.198 & 0.980 \\
$MLN$ & 250  & 0.014 & 0.038 & 0.059 & 0.029 & 0.088 & 0.106 & 0.091 & 0.805 \\
      & 500  & 0.042 & 0.047 & 0.054 & 0.041 & 0.109 & 0.199 & 0.182 & 0.995 \\
      & 750  & 0.055 & 0.059 & 0.070 & 0.050 & 0.123 & 0.269 & 0.228 & 1.000 \\
      & 1000 & 0.066 & 0.083 & 0.063 & 0.082 & 0.253 & 0.462 & 0.421 & 1.000 \\
      \bottomrule
\end{tabular}%
}
\label{Table Power DGP3}%
\end{table}

\begin{table}[!htbp] \centering\renewcommand{\arraystretch}{1.15}%
\addtolength{\tabcolsep}{12.0pt}
\caption{Empirical rejection probabilities of the tests under DGPP.4.}%
\setlength{\tabcolsep}{3mm}{
\begin{tabular}
[c]{cccccccccc}\toprule
Error & $T$   & $\widehat{\mathcal{D}}$ & $\widehat{\mathcal{D}}_{b}$ & $MS_{A}$ &
$MQ_{A}$ & $MS_M^*$ & $MQ_M^*$ & $T_{CM}^*$ & $\widehat{U}_{T}$ \\ \midrule
$MN$  & 250  & 0.450 & 0.473 & 0.066 & 0.198 & 0.048 & 0.044 & 0.045 & 0.515 \\
      & 500  & 0.759 & 0.771 & 0.111 & 0.262 & 0.065 & 0.068 & 0.067 & 0.849 \\
      & 750  & 0.944 & 0.933 & 0.168 & 0.398 & 0.065 & 0.105 & 0.091 & 0.989 \\
      & 1000 & 0.989 & 0.981 & 0.256 & 0.498 & 0.121 & 0.159 & 0.140 & 0.998 \\
$MT$  & 250  & 0.134 & 0.149 & 0.050 & 0.121 & 0.028 & 0.027 & 0.027 & 0.217 \\
      & 500  & 0.130 & 0.162 & 0.091 & 0.157 & 0.067 & 0.068 & 0.068 & 0.515 \\
      & 750  & 0.174 & 0.194 & 0.110 & 0.147 & 0.078 & 0.087 & 0.084 & 0.825 \\
      & 1000 & 0.214 & 0.212 & 0.134 & 0.213 & 0.095 & 0.129 & 0.121 & 0.940 \\
$SMT$ & 250  & 0.136 & 0.160 & 0.060 & 0.128 & 0.030 & 0.038 & 0.037 & 0.272 \\
      & 500  & 0.228 & 0.254 & 0.080 & 0.182 & 0.054 & 0.062 & 0.058 & 0.641 \\
      & 750  & 0.352 & 0.313 & 0.106 & 0.180 & 0.094 & 0.096 & 0.095 & 0.879 \\
      & 1000 & 0.395 & 0.366 & 0.121 & 0.188 & 0.139 & 0.140 & 0.140 & 0.964 \\
$MLN$ & 250  & 0.070 & 0.123 & 0.070 & 0.127 & 0.043 & 0.038 & 0.039 & 0.806 \\
      & 500  & 0.117 & 0.128 & 0.092 & 0.180 & 0.060 & 0.087 & 0.084 & 0.985 \\
      & 750  & 0.139 & 0.136 & 0.108 & 0.181 & 0.104 & 0.132 & 0.116 & 1.000 \\
      & 1000 & 0.139 & 0.156 & 0.141 & 0.247 & 0.242 & 0.328 & 0.305 & 1.000 \\
      \bottomrule
\end{tabular}%
}
\label{Table Power DGP4}%
\end{table}

We next consider the multiple-break designs. DGPP.2 contains two abrupt breaks, and $\widehat U_T$ records the highest rejection frequencies under both Gaussian and non-Gaussian innovations. By comparison, $MS_A$ and $MQ_A$, which are designed primarily for single-break alternatives, have relatively low power in this setting. The LAD-based procedures $MS_M^*$, $MQ_M^*$, and $T_{CM}^*$ perform better under some non-Gaussian distributions, but their rejection frequencies remain below those of $\widehat U_T$ in the reported experiments. The gains from $\widehat U_T$ are consistent with its aggregation of score information across multiple quantile levels. DGPP.3, which contains four abrupt breaks, produces a similar ranking of the tests. Taken together, these results indicate that $\widehat U_T$ retains power when volatility changes occur at several points in the sample.

\begin{table}[!htbp] \centering\renewcommand{\arraystretch}{1.15}%
\addtolength{\tabcolsep}{12.0pt}
\caption{Empirical rejection probabilities of the tests under DGPP.5.}%
\setlength{\tabcolsep}{3mm}{
\begin{tabular}
[c]{cccccccccc}\toprule
Error & $T$   & $\widehat{\mathcal{D}}$ & $\widehat{\mathcal{D}}_{b}$ & $MS_{A}$ &
$MQ_{A}$ & $MS_M^*$ & $MQ_M^*$ & $T_{CM}^*$ & $\widehat{U}_{T}$ \\ \midrule
$MN$  & 250  & 0.435 & 0.464 & 0.365 & 0.418 & 0.239 & 0.257 & 0.254 & 0.358 \\
      & 500  & 0.797 & 0.821 & 0.738 & 0.805 & 0.588 & 0.610 & 0.599 & 0.708 \\
      & 750  & 0.959 & 0.953 & 0.923 & 0.972 & 0.727 & 0.805 & 0.791 & 0.969 \\
      & 1000 & 0.993 & 0.986 & 0.984 & 0.997 & 0.914 & 0.939 & 0.927 & 0.998 \\
$MT$  & 250  & 0.104 & 0.121 & 0.139 & 0.167 & 0.163 & 0.176 & 0.176 & 0.161 \\
      & 500  & 0.112 & 0.137 & 0.280 & 0.266 & 0.412 & 0.459 & 0.454 & 0.372 \\
      & 750  & 0.156 & 0.181 & 0.404 & 0.357 & 0.585 & 0.645 & 0.633 & 0.711 \\
      & 1000 & 0.199 & 0.200 & 0.495 & 0.519 & 0.822 & 0.862 & 0.858 & 0.893 \\
$SMT$ & 250  & 0.095 & 0.125 & 0.144 & 0.154 & 0.167 & 0.191 & 0.183 & 0.187 \\
      & 500  & 0.196 & 0.231 & 0.273 & 0.317 & 0.461 & 0.477 & 0.469 & 0.488 \\
      & 750  & 0.336 & 0.297 & 0.468 & 0.453 & 0.733 & 0.755 & 0.746 & 0.793 \\
      & 1000 & 0.399 & 0.371 & 0.554 & 0.562 & 0.869 & 0.887 & 0.886 & 0.943 \\
$MLN$ & 250  & 0.053 & 0.096 & 0.269 & 0.273 & 0.474 & 0.498 & 0.495 & 0.678 \\
      & 500  & 0.088 & 0.101 & 0.520 & 0.554 & 0.865 & 0.911 & 0.906 & 0.969 \\
      & 750  & 0.112 & 0.120 & 0.673 & 0.704 & 0.977 & 0.984 & 0.981 & 1.000 \\
      & 1000 & 0.114 & 0.125 & 0.783 & 0.859 & 1.000 & 1.000 & 1.000 & 1.000 \\ \bottomrule
\end{tabular}%
}
\label{Table Power DGP5}%
\end{table}

The remaining designs consider smooth changes in volatility. DGPP.4 specifies a symmetric U-shaped path. With Gaussian innovations, the rejection frequencies of $\widehat U_T$ and $\widehat{\mathcal D}$ are close, with a small advantage for $\widehat U_T$. Under the non-Gaussian distributions, $\widehat U_T$ has the highest rejection frequencies, and the differences from the benchmark tests are larger. DGPP.5 uses a quadratic volatility path. In this case, $\widehat{\mathcal D}$ performs slightly better under Gaussian innovations, whereas $\widehat U_T$ records higher rejection frequencies under the non-Gaussian designs. For the oscillatory path in DGPP.6, $\widehat U_T$ has the highest rejection frequency for each innovation distribution considered.

\begin{table}[!htbp] \centering\renewcommand{\arraystretch}{1.15}%
\addtolength{\tabcolsep}{12.0pt}
\caption{Empirical rejection probabilities of the tests under DGPP.6.}%
\setlength{\tabcolsep}{3mm}{
\begin{tabular}
[c]{cccccccccc}
\toprule
Error & $T$   & $\widehat{\mathcal{D}}$ & $\widehat{\mathcal{D}}_{b}$ & $MS_{A}$ &
$MQ_{A}$ & $MS_M^*$ & $MQ_M^*$ & $T_{CM}^*$ & $\widehat{U}_{T}$ \\ \midrule
$MN$  & 250  & 0.574 & 0.558 & 0.720 & 0.547 & 0.447 & 0.401 & 0.425 & 0.944 \\
      & 500  & 0.977 & 0.950 & 0.979 & 0.915 & 0.910 & 0.852 & 0.883 & 0.999 \\
      & 750  & 0.997 & 0.993 & 1.000 & 0.997 & 0.989 & 0.973 & 0.984 & 1.000 \\
      & 1000 & 1.000 & 0.998 & 1.000 & 1.000 & 1.000 & 0.995 & 1.000 & 1.000 \\
$MT$  & 250  & 0.099 & 0.117 & 0.212 & 0.175 & 0.272 & 0.253 & 0.261 & 0.641 \\
      & 500  & 0.168 & 0.222 & 0.459 & 0.333 & 0.750 & 0.670 & 0.709 & 0.956 \\
      & 750  & 0.283 & 0.308 & 0.624 & 0.440 & 0.915 & 0.862 & 0.898 & 1.000 \\
      & 1000 & 0.370 & 0.362 & 0.730 & 0.605 & 0.984 & 0.971 & 0.981 & 1.000 \\
$SMT$ & 250  & 0.116 & 0.158 & 0.281 & 0.185 & 0.347 & 0.316 & 0.320 & 0.768 \\
      & 500  & 0.358 & 0.364 & 0.540 & 0.430 & 0.791 & 0.737 & 0.774 & 0.982 \\
      & 750  & 0.555 & 0.502 & 0.734 & 0.570 & 0.962 & 0.923 & 0.944 & 1.000 \\
      & 1000 & 0.655 & 0.597 & 0.814 & 0.675 & 0.993 & 0.978 & 0.988 & 1.000 \\
$MLN$ & 250  & 0.058 & 0.104 & 0.419 & 0.280 & 0.769 & 0.712 & 0.739 & 0.997 \\
      & 500  & 0.117 & 0.140 & 0.625 & 0.491 & 0.996 & 0.981 & 0.992 & 1.000 \\
      & 750  & 0.197 & 0.211 & 0.795 & 0.620 & 1.000 & 1.000 & 1.000 & 1.000 \\
      & 1000 & 0.224 & 0.243 & 0.885 & 0.790 & 1.000 & 1.000 & 1.000 & 1.000 \\ \bottomrule
\end{tabular}%
}
\label{Table Power DGP6}%
\end{table}

\begin{figure}[h!]
\begin{center}
\includegraphics[height=8.018in,width=7.0553in, trim=1.5cm 0cm 0cm 0cm]{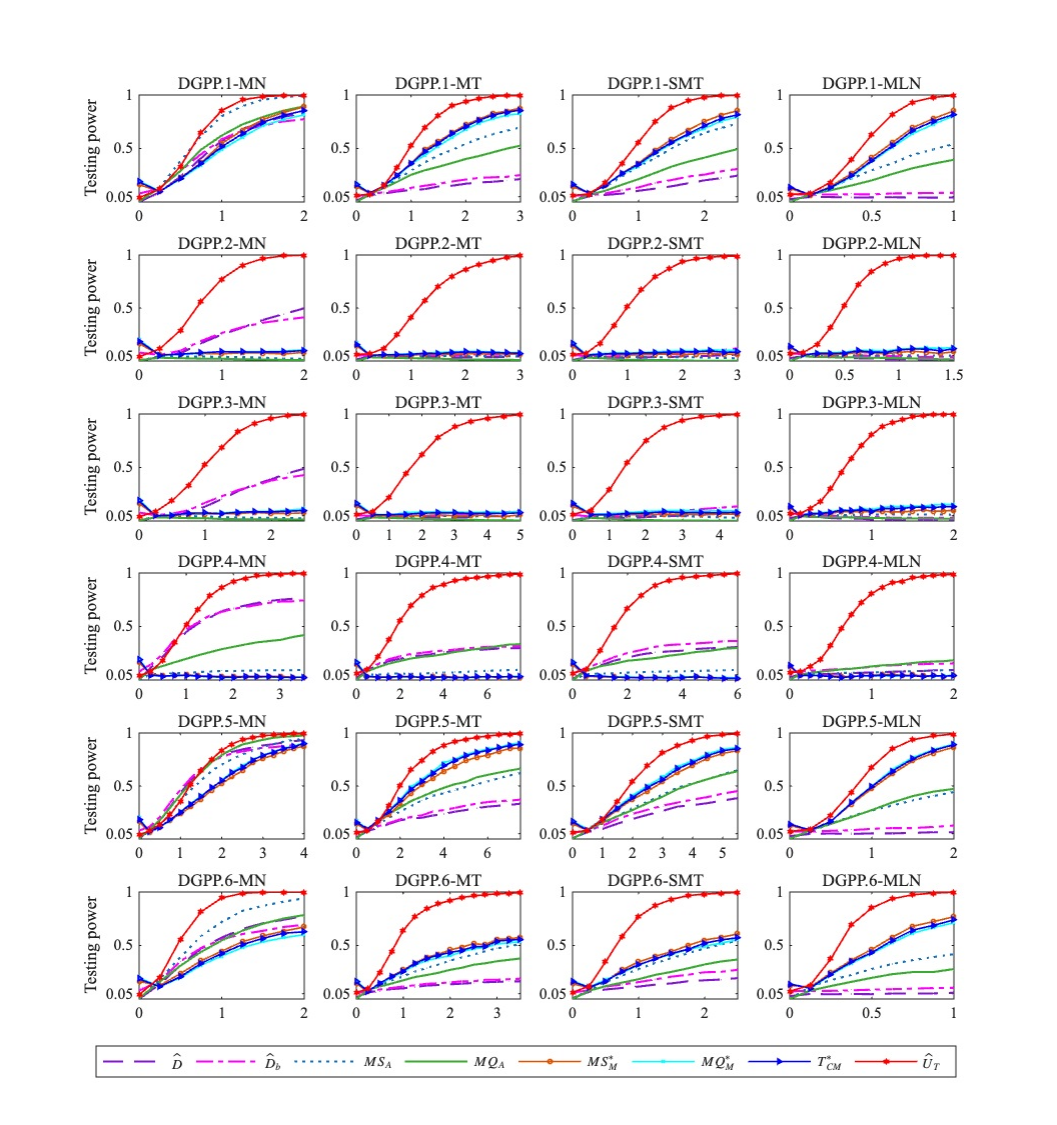}%
\caption{Empirical power curves of all tests under DGPP.1-6 with different distributions.}%
\label{figure 1}
\end{center}
\end{figure}

Figure~\ref{figure 1} examines how the rejection frequencies vary with the magnitude of the departure, $\gamma$, in DGPP.1--6. All other parameters are held fixed and $T=250$; the results for $T=500$, $750$, and $1000$ are similar. The null corresponds to $\gamma=0$. The rows of the figure index the innovation distributions, and the columns index the six volatility paths. For $\widehat U_T$, the rejection frequency increases monotonically over the reported grid of $\gamma$ in every design. The increase is relatively steep in most of the multiple-break and smooth-change experiments. Some benchmark procedures respond more slowly to the departure, and several of their simulated power curves are not monotone over the same grid. In a few designs, their rejection frequencies remain close to the nominal level even as $\gamma$ increases.

As an additional robustness check, we examine whether the finite-sample performance of the proposed tests is affected when the innovation process is not directly observed. This setting is relevant in empirical applications, where volatility tests are often applied to residuals obtained after filtering conditional mean dynamics. To this end, we generate the observed process from the five-dimensional VAR(2) model
\begin{align}
Y_t = \mathcal A_1Y_{t-1}+\mathcal A_2Y_{t-2}+u_t, \qquad t=1,\ldots,T,
\label{VAR2}
\end{align}
where $Y_t=(Y_{1t},\ldots,Y_{5t})^\top$ and $u_t=(u_{1t},\ldots,u_{5t})^\top$. The innovation $u_t$ is generated according to the same designs as in \eqref{SDGP}. The coefficient matrix $\mathcal A_1$ is a symmetric Toeplitz matrix with $(i,j)$th element $0.3(0.5)^{|i-j|}$, $i,j=1,\ldots,5$, and $\mathcal A_2=0.15I_5$.

In each replication, the conditional mean is first estimated by a VAR model using least squares, and the resulting residuals $\hat u_t$ are used in place of $u_t$ to construct the proposed tests. To avoid imposing the true lag order, the VAR lag length is selected by the corrected portmanteau procedure of \cite{patilea2013corrected}, which is designed to be robust to time-varying unconditional variances. The corresponding results, reported in the supplementary material, show that the residual-based implementation delivers size and power close to those obtained when the innovations are observed. Hence, preliminary estimation of the conditional mean has little finite-sample impact on the proposed testing procedure.

\bigskip

\section{Empirical Application}

Factor models are widely used in financial econometrics to represent the comovement of a large number of asset returns through a small set of common components \citep{ross1976arbitrage}. This parsimonious structure alleviates the dimensionality problems encountered in covariance estimation and asset-pricing applications \citep{bai2002determining,fan2013large}, and has motivated the development of a broad range of observed and latent pricing factors \citep{feng2020taming}. Many factor-based procedures nevertheless treat the covariance matrix of the common factors as constant over the sample period \citep{fan2011high,giglio2021asset,caner2023sharpe,fan2022time}. This restriction may be consequential when the sample spans episodes of substantial financial and economic disruption. Assessing factor-covariance stability is therefore relevant before imposing a time-invariant covariance structure in subsequent empirical analysis.

We examine this issue using the Fama--French three-factor model of \citet{fama1993common}. The model comprises the market factor (MKT), the size factor (SMB), and the value factor (HML), and remains a standard benchmark in empirical asset pricing. Because the covariance matrix of these factors summarizes their joint variation and enters many risk-management and portfolio applications, the three-factor model provides a natural setting for the proposed structural-change test.

\begin{figure}[h!]
		\begin{center}
\includegraphics[height=2in,width=6.8553in, trim=1cm 0cm -1cm 0cm]{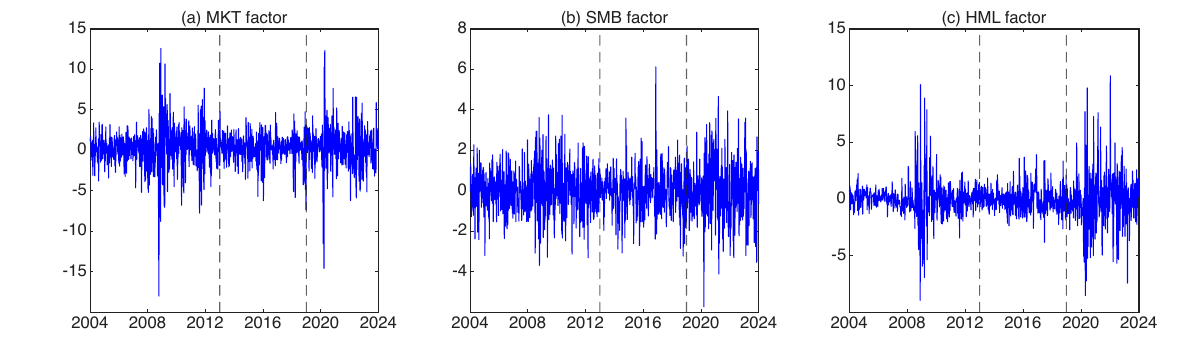}%
\caption{Time series plots of the weekly Fama-French three factors.}%
\label{figure 2}
\end{center}
	\end{figure}

Weekly factor returns are obtained from the Ken French Data Library\footnote{\url{https://mba.tuck.dartmouth.edu/pages/faculty/ken.french/data_library.html}.} for the period from January 2, 2004, to December 29, 2023. We consider the full sample and three subsamples, 2004--2012, 2013--2018, and 2019--2023, which encompass the Global Financial Crisis, the Brexit referendum, and the COVID-19 pandemic, respectively. Figure~\ref{figure 2} plots the three factor returns. Their variability changes considerably over the sample, with particularly pronounced fluctuations during the 2008 financial crisis and the onset of the COVID-19 pandemic in 2020. Quieter intervals are also evident, especially during parts of the middle subsample. These differences in scale and comovement motivate a formal examination of whether the factor covariance matrix remains constant over the full sample and within the three subsamples.

\begin{figure}[h!]
\begin{center}
\includegraphics[height=2.0in,width=6.8553in, trim=1.6cm 0cm 0cm 0cm]{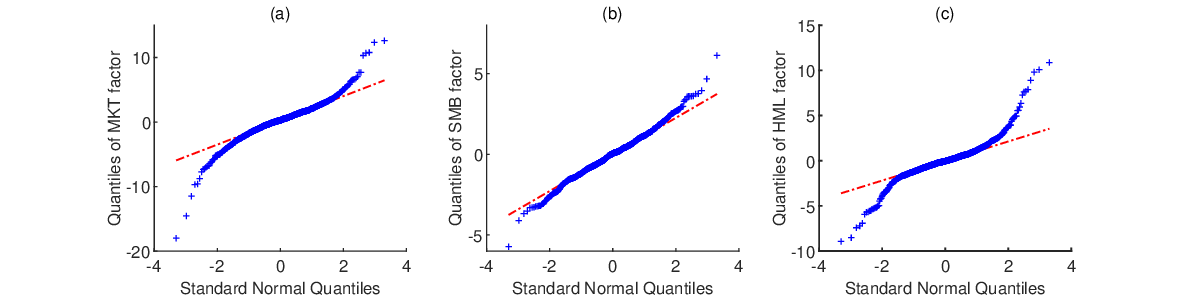}%
\caption{Q-Q plots of Fama-French three factors.}%
\label{figure 3}
\end{center}
\end{figure}

	Figure \ref{figure 3} further presents the Q-Q plots for the three factors against the standard normal distribution. Notable deviations from normality are observed, particularly for MKT and HML, indicating that the distributions of these factors are characterized by heavy tails.  Table \ref{Descriptive statistics} reports the descriptive statistics for the Fama-French three factors across the full periods and three  sub-periods. The MKT and HML factors exhibit high leptokurtic and heavy-tailed distributions throughout all periods, as evidenced by Kurtosis values well exceeding 3 and the rejection of normality by the Jarque-Bera (JB) test ($p < 0.01$). Similarly, the SMB factor displays leptokurtic features in most periods, although its distribution appears relatively closer to normal during the 2004-2012 and 2019-2023 sub-periods, where the JB test fails to reject the null hypothesis at the 5\% level. These pervasive heavy-tailed features across the factors justify the use of quantile-based testing frameworks, which provide greater robustness to non-normal errors than traditional mean-variance approaches. Consequently, we anticipate that our new test will demonstrate enhanced effectiveness in detecting structural changes in the covariance matrix of the Fama-French three factors.

    \begin{table}[!h] \centering\renewcommand{\arraystretch}{1.5}%
		\addtolength{\tabcolsep}{15.0pt}
        \begin{threeparttable}
		\caption{Descriptive statistics of the Fama-French three factors}
		\setlength{\tabcolsep}{3.75mm}{
			\begin{tabular}
				[c]{cccccc}\toprule
				& Factor & $SD$ & $Skewness$ &
				$Kurtosis$ & $JB\,\,test$ \\ 
                \midrule
				$\hspace{1.6mm}2004-2023\,(T=1044)$& $MKT$ & $2.486$ & $-0.550$ & $9.486$ & $0.001$ \\
                & $SMB$ & $1.263$ & $0.105$ & $4.217$ & $0.001$\\
                & $HML$ & $1.757$ & $0.475$ & $10.407$ & $0.001$\\
				$2004-2012\,(T=470)$& $MKT$ & $2.672$ & $-0.509$ & $9.488$ & $0.001$ \\
                & $SMB$ & $1.205$ & $0.021$ & $3.355$ & $0.259$\\
                & $HML$ & $1.544$ & $0.692$ & $14.605$ & $0.001$\\
				$2013-2018\,(T=313)$& $MKT$ & $1.779$ & $-0.759$ & $5.074$ & $0.001$ \\ 
                & $SMB$ & $1.087$ & $0.368$ & $6.267$ & $0.001$\\
                & $HML$ & $1.107$ & $0.561$ & $4.362$ & $0.001$\\
				$2019-2023\,(T=261)$& $MKT$ & $2.847$ & $-0.475$ & $7.952$ & $0.001$ \\
                & $SMB$ & $1.535$ & $0.071$ & $3.579$ & $0.115$\\
                & $HML$ & $2.571$ & $0.279$ & $5.510$ & $0.001$\\
                \bottomrule
			\end{tabular}%
		}
		\label{Descriptive statistics}%
        \begin{tablenotes}
            \footnotesize
            \item \textit{Notes}: The SD, skewness, and kurtosis are reported as their sample estimates, while the JB test is reported in terms of its p-value.
        \end{tablenotes}
        \end{threeparttable}
	\end{table}%

Table~\ref{Table emp1} reports the $p$-values for tests of stability in the covariance matrix of the Fama--French three factors. Over the full sample, 2004--2023, $\widehat U_T$, $MS_A$, $MQ_A$, and the three LAD-based procedures reject the null at the 1\% level. The corresponding $p$-values for $\widehat{\mathcal D}$ and $\widehat{\mathcal D}_b$ are 0.097 and 0.154, respectively. The subsample results differ more across procedures. For 2004--2012 and 2013--2018, $\widehat U_T$ and the three LAD-based tests reject at the 1\% level. During 2019--2023, however, the LAD-based tests do not reject, whereas the $p$-value of $\widehat U_T$ remains below 0.001. Among the remaining procedures, $\widehat{\mathcal D}_b$ rejects at the 10\% level for 2004--2012, and $MQ_A$ rejects at the 10\% level for 2013--2018; neither provides evidence against stability during 2019--2023. Thus, in this application, $\widehat U_T$ is the only procedure that rejects covariance stability at the 1\% level in the full sample and in each subsample. This comparison concerns the reported data and should not be interpreted as a general ranking of test power.

	\begin{table}[!h] \centering\renewcommand{\arraystretch}{1.5}%
		\addtolength{\tabcolsep}{15.0pt}
        \begin{threeparttable}
		\caption{The testing results for Fama-French three factors}
		\setlength{\tabcolsep}{2.75mm}{
			\begin{tabular}
				[c]{ccccccccc}\toprule
				& $\hat{\mathcal{D}}$ & $\hat{\mathcal{D}}_{b}$ & $MS_{A}$ &
				$MQ_{A}$ & $MS_M^*$ & $MQ_M^*$ & $T_{CM}^*$ & $\widehat{U}_{T}$\\ \midrule
				$\hspace{1.6mm}2004-2023\,(T=1044)$& $0.097$ & $0.154$ & $0.000$ & $0.000$ & $0.000$ & $0.000$ & $0.000$ & $0.000$\\
				$2004-2012\,(T=470)$& $0.125$ & $0.097$ & $0.133$ & $0.274$ & $0.000$ & $0.000$ & $0.000$ & $0.000$\\
				$2013-2018\,(T=313)$& $0.222$ & $0.110$ & $0.380$ & $0.071$ & $0.001$ & $0.000$ & $0.000$ & $0.000$\\ 
				$2019-2023\,(T=261)$& $0.511$ & $0.411$ & $0.135$ & $0.102$ & $0.306$ & $0.194$ & $0.240$ & $0.000$\\ \bottomrule
			\end{tabular}%
		}
		\label{Table emp1}%
        \begin{tablenotes}
            \footnotesize
            \item \textit{Notes}: The results in the table are the p-values of each test statistic.
        \end{tablenotes}
        \end{threeparttable}
	\end{table}%

\begin{figure}[h!]
		\begin{center}
			\includegraphics[height=3.6in,width=6.1553in, trim=0cm 0cm 0cm 0cm]{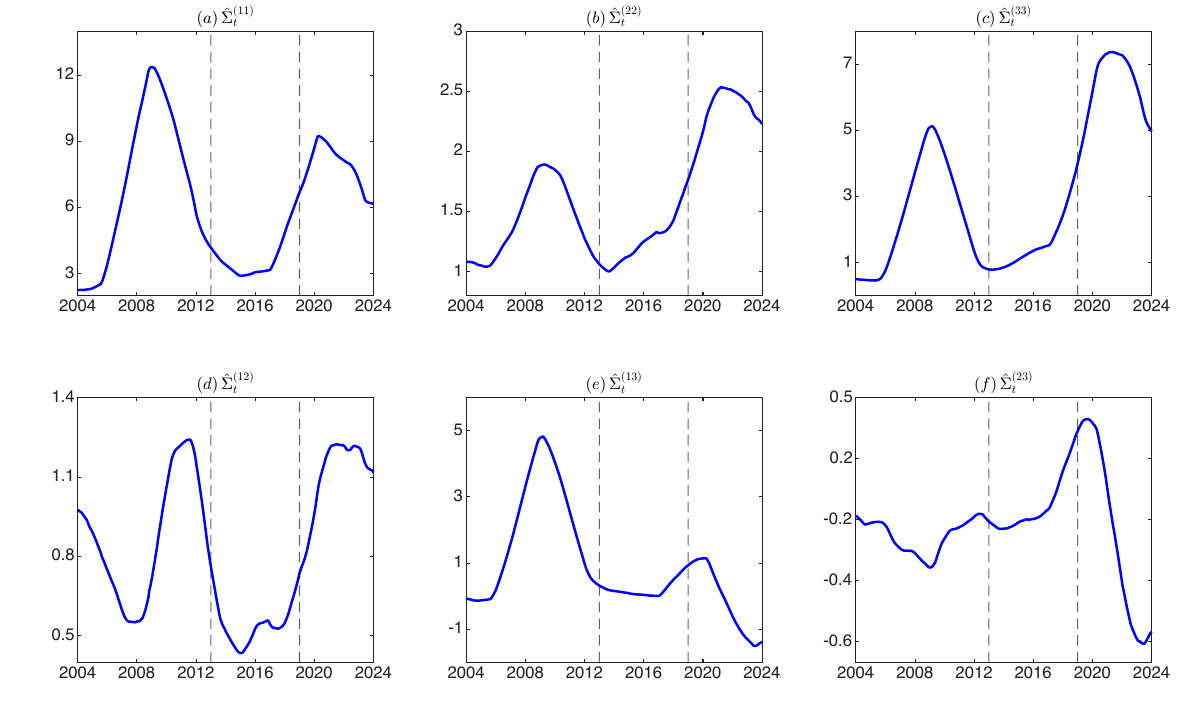}%
			\caption{The nonparametric estimate of ${\Sigma}_t$.}%
			\label{figure 4}
		\end{center}
	\end{figure}

   To further characterize the temporal evolution of the second-moment dynamics, we employ a local constant kernel estimator to obtain the nonparametric path of the covariance matrix $\Sigma_t$.\footnote{Let $\{u_t\}_{t=1}^T$ denote the sequence of factor innovations. The local constant estimator is defined as $\hat{\Sigma}_t = \left( \sum_{s=1}^T k_{t,s} \right)^{-1} \sum_{s=1}^T k_{t,s} u_s u_s^{\intercal}$, where $k_{t,s} = k((t-s)/Th)$ denotes the kernel weight, with $k(\cdot)$ representing the Bartlett kernel and $h$ the bandwidth parameter. The optimal bandwidth $h$ is selected via leave-one-out cross-validation to minimize the integrated mean squared error.} Figure \ref{figure 4} illustrates these estimates, with subplots (a)-(c) depicting individual factor variances and subplots (d)-(f) displaying the pairwise covariances. The estimated trajectories exhibit pronounced time-variation, particularly around major market disruptions. These visual dynamics reinforce our formal testing results, confirming that the Fama-French factor covariance matrix is subject to significant structural instability rather than remaining constant over the sample period.

The results provide evidence against stability of the Fama--French factor covariance matrix, both over the full sample and within the three shorter periods. For this application, treating the covariance matrix as constant over 2004--2023 therefore appears restrictive and may affect subsequent risk or portfolio analysis. The local covariance estimates are included only as descriptive diagnostics and do not enter the test statistic. They indicate which components may contribute to rejection, but they neither estimate break dates consistently nor imply that all variances and covariances change at the same time. Componentwise plots and more formal post-rejection procedures would be needed to identify the timing and sources of the detected instability.

\bigskip

\section{Conclusion}

This paper develops a multiple-quantile test for structural instability in multivariate volatility. The test aggregates bounded quantile scores and admits a weighted leave-$q$-out $U$-statistic representation. Excluding nearby index pairs makes the centering effect induced by serial dependence asymptotically negligible, while the remaining nuisance quantities are estimated under the null. After feasible variance standardization, the statistic has a standard normal null limit. The bounded-score construction avoids the finite fourth- or eighth-moment conditions commonly imposed by least-squares and quasi-likelihood procedures. The test is consistent against fixed alternatives with a positive integrated quantile-score signal and has nontrivial local power against both smooth departures and sharp transitions approaching multiple structural breaks. The simulations show satisfactory size and favorable power under heavy-tailed innovations, with competitive performance under Gaussian innovations. The empirical application provides evidence against stability of the Fama--French factor covariance matrix over the full sample and several economically relevant subsamples. Two extensions warrant further study. The first allows the dimension of the volatility matrix to increase with the sample size, which would require regularized estimation of the quantile matrices and their long-run cross-covariance structure. The second concerns post-rejection inference on the locations, magnitudes, and temporal profiles of volatility changes. Such methods would help identify the components and periods responsible for rejection of the omnibus null.

\bibliographystyle{chicago}
\bibliography{reference}
\end{document}